\documentclass[aps, prd, twocolumn, preprintnumbers, superscriptaddress, nofootinbib, floatfix]{revtex4-1}

\usepackage{amsmath}	
\usepackage{amsthm}		
\usepackage{amssymb}	
\usepackage{amsfonts}
\usepackage{eufrak}
\usepackage{datetime}
\usepackage{graphicx}
\usepackage{color}
\usepackage{verbatim}
\usepackage{bm}
\usepackage[colorlinks=true, citecolor=midblue, linkcolor=midblue, urlcolor=midblue]{hyperref}
\usepackage{mciteplus}
\usepackage{epigraph}
\usepackage{array}
\usepackage{soul}

\usepackage{setspace}

\definecolor{grey}{rgb}{0.4,0.4,0.4}
\definecolor{dullmagenta}{rgb}{0.4,0,0.4}
\definecolor{darkblue}{rgb}{0,0,0.4}
\definecolor{midblue}{rgb}{0,0,0.7}
\definecolor{midred}{rgb}{0.5,0,0}
\definecolor{orange}{rgb}{1,0.5,0}
\definecolor{lightbrown}{rgb}{0.75,0.5,0.25}
\definecolor{tan}{cmyk}{0.14,0.42,0.56,0}
\definecolor{djunglegreen}{cmyk}{0.99,0,0.52,0}
\definecolor{lightgreen}{rgb}{0,1,0}
\definecolor{olivegreen}{cmyk}{0.64,0,0.95,0.40}
\definecolor{midgreen}{rgb}{0.0,0.675,0.0}
\definecolor{darkgreen}{rgb}{0,0.5,0}

\newcommand{\qq}{\qquad}

\newcommand{\vs}{\vspace}

\renewcommand{\.}{\hspace{0.5mm}}

\newcommand{\la}{\ensuremath{\leftarrow}}

\newcommand{\Brm}{\ensuremath{\mathrm{B}}}
\newcommand{\Crm}{\ensuremath{\mathrm{C}}}

\newcommand{\Erm}{\ensuremath{\mathrm{E}}}
\newcommand{\Frm}{\ensuremath{\mathrm{F}}}

\newcommand{\Krm}{\ensuremath{\mathrm{K}}}
\newcommand{\Lrm}{\ensuremath{\mathrm{L}}}

\newcommand{\Rrm}{\ensuremath{\mathrm{R}}}
\newcommand{\Srm}{\ensuremath{\mathrm{S}}}
\newcommand{\Trm}{\ensuremath{\mathrm{T}}}

\newcommand{\arm}{\ensuremath{\mathrm{a}}}

\newcommand{\grm}{\ensuremath{\mathrm{g}}}

\newcommand{\mrm}{\ensuremath{\mathrm{m}}}

\newcommand{\orm}{\ensuremath{\mathrm{o}}}
\newcommand{\prm}{\ensuremath{\mathrm{p}}}

\newcommand{\srm}{\ensuremath{\mathrm{s}}}

\newcommand{\Ocal}{\ensuremath{\mathcal{O}}}

\renewcommand{\d}{\ensuremath{\mathrm{d}}}

\newcommand{\ie}{i.e.}

\newcommand{\be}{\begin{equation}}
\newcommand{\ee}{\end{equation}}
\newcommand{\ba}{\begin{eqnarray}}
\newcommand{\ea}{\end{eqnarray}}

\smallskipamount

\settimeformat{ampmtime}

\usepackage{float}

\def\ga{\mathrel{\raise.3ex\hbox{$>$\kern-.75em\lower1ex\hbox{$\sim$}}}}
\def\la{\mathrel{\raise.3ex\hbox{$<$\kern-.75em\lower1ex\hbox{$\sim$}}}}

\def\Msun{M_\odot}

\def\fPBH{f_{\rm PBH}}

\def\fPBH{f_{\rm PBH}}

\newcommand{\exclude}[1]{}

\enlargethispage{\baselineskip}

\hypersetup{
	colorlinks = true,
	citecolor = blue
}

\everymath{\displaystyle} 
\newcommand{\MP}{M_{\rm Pl}}
\newcommand{\zeq}{z_{\rm eq}}
\newcommand{\sv}{\langle\sigma v \rangle}

\renewcommand\({\left(}
\renewcommand\){\right)}
\renewcommand\[{\left[}
\renewcommand\]{\right]}

\begin{document}

\title{Constraints on Stupendously Large Black Holes}

\newcommand{\FIRSTAFF}{
	\affiliation{
		School of Physics and Astronomy, 
		Queen Mary University of London,
		Mile End Road, 
		London E1 4NS, 
		UK}}

\newcommand{\SECONDAFF}{
	\affiliation{
		Research Center for the Early Universe, 
		University of Tokyo, 
		Tokyo 113-0033, 
		Japan}}

\newcommand{\THIRDAFF}{
	\affiliation{
		Gravitation Astroparticle Physics Amsterdam (GRAPPA),\\
		Institute for Theoretical Physics Amsterdam and Delta Institute for Theoretical Physics,\\
		University of Amsterdam, Science Park 904, 1098 XH Amsterdam, The Netherlands}}

\newcommand{\FOURTHAFF}{
	\affiliation{
		Arnold Sommerfeld Center,
		Ludwig-Maximilians-Universit{\"a}t,
		Theresienstra{\ss}e 37,
		80333 M{\"u}nchen,
		Germany}}
			
\author{Bernard Carr}
	\email[Electronic address: ]{B.J.Carr@qmul.ac.uk}
	\FIRSTAFF
	\SECONDAFF

\author{Florian K{\"u}hnel}
	\email[Electronic address: ]{kuhnel@kth.se}
	\FOURTHAFF

\author{Luca Visinelli}
	\email[Electronic address: ]{l.visinelli@uva.nl}
	\THIRDAFF

\date{\formatdate{\day}{\month}{\year}, \currenttime}

\begin{abstract}
\vs{-3mm}We consider the observational constraints on stupendously large black holes (SLABs) in the mass range $M \gtrsim 10^{11}\,M_{\odot}$. These have attracted little attention hitherto and we are aware of no published constraints on a SLAB population in the range $(10^{12}$ -- $10^{18})\,M_{\odot}$. However, there is already evidence for black holes of up to nearly $10^{11}\,M_{\odot}$ in galactic nuclei, so it is conceivable that SLABs exist and they may even have been seeded by primordial black holes. We focus on limits associated with 
	(i) dynamical and lensing effects, 
	(ii) the generation of background radiation through the accretion of gas during the pregalactic epoch, and 
	(iii) the gamma-ray emission from the annihilation of the halo of weakly interacting massive particles (WIMPs) expected to form around each SLAB if these provide the dark matter.
Finally, we comment on the constraints on the mass of ultra-light bosons from future measurements of the mass and spin of SLABs.
\end{abstract}

\maketitle

\section{Introduction}
\label{sec:Introduction}

Black holes (BHs) are a key prediction of general relativity. There are a plethora of observations indicating their existence in the solar~\cite{2006ARA&A..44...49R} or intermediate-mass~\cite{Lin:2020exl} range. In particular, the existence of binary black holes in the mass range $( 10$ -- $50 )\,M_{\odot}$ has been demonstrated by the detection of gravitational waves from inspiralling binaries~\cite{LIGOScientific:2018mvr}.

There is also evidence for supermassive black holes (SMBHs) at the centres of galaxies~\cite{Kormendy:1995er}, including Sagittarius A$^{*}$ at the centre of our own galaxy, with a mass of $4 \times 10^{6}\,M_{\odot}$~\cite{Schodel:2002vg}. Recently, the imaging of the shadow created by M87$^{*}$, the SMBH at the centre of the giant elliptical galaxy M87 with a mass of $6.5 \times 10^{9}\,M_{\odot}$, has been reported by the Event Horizon Telescope~\cite{Akiyama:2019cqa}. The SMBHs in galactic nuclei span a huge mass range, extending up to nearly $10^{11}\,M_{\odot}$~\cite{2011Natur.480..215M}. The current heaviest BH is associated with the quasar TON 618 and has a mass of $7 \times 10^{10}\,M_{\odot}$~\cite{Shemmer:2004ph}, while the second heaviest, at the centre of the galaxy IC 1101, has a mass inferred from its radio emission of $4 \times 10^{10}\,M_{\odot}$~\cite{2017MNRAS.471.2321D}.

This raises the issue of whether there could be even larger BHs in galactic nuclei and whether indeed there is any natural upper limit to the mass of a SMBH. We will not review the extensive literature on this topic here but this can be found in the pioneering paper of Natarajan and Treister~\cite{Natarajan:2008ks}, who investigated the possibility of ``ultra-massive'' black holes, defined to be ones larger than $5 \times 10^{9}\,M_{\odot}$. They argued that their existence is implied by the observed relationship between the SMBH mass and the bulge luminosity of the host galaxy. It is also a natural implication of the accretion and merger of smaller SMBHs, so one expects a few such objects in central galaxies in clusters. However, they concluded on the basis of observational and theoretical considerations and the known local SMBH mass function that there is a natural upper limit on the mass of $\sim 10^{11}\,M_{\odot}$.

More recently, based on the assumption that SMBHs are hosted by disc galaxies, King~\cite{10.1093/mnrasl/slv186} has derived an upper limit of $5 \times 10^{10}\,M_{\odot}$ (or $3 \times 10^{11}\,M_{\odot}$ for maximal prograde spin) above which they cannot grow through luminous accretion of gas, the precise value depending on the properties of the host galaxy. He points out that the associated Eddington luminosity is close to the largest observed AGN bolometric luminosity. Black holes can still grow above this mass by non-luminous means, such as mergers, but they cannot become luminous again. A related upper limit has been obtained by Yazdi and Afshordi~\cite{Yazdi:2016fyt}. Both arguments are based on the Toomre instability of $\alpha$-disks~\cite{Toomre:1964zx}: if the SMBH is too large, it ceases to grow through accretion because the disk becomes too massive and fragments under self-gravity. This leads to a strict upper limit on the mass of SMBHs as a function of cosmic time and spin.

We describe these models in more detail in Sec.~\ref{sec:Formation-of-SLABs}. All the estimates of the maximum SMBH mass are in rough agreement with the observations but they are dependent on details of the accretion models. They also leave open the possibility that more massive SMBHs could {\it exist} even if they cannot be luminous and this provides the motivation for the present paper.
 
We will describe BHs larger than $10^{11}\,M_{\odot}$ (\ie~larger than the SMBHs currently observed in galactic nuclei) as ``Stupendously Large Black Holes'' or ``SLABs'', this extending well beyond the ultra-massive range of Ref.~\cite{Natarajan:2008ks}. While the most natural assumption is that SLABs represent the high-mass tail of the population of SMBHs in galactic nuclei, we also consider the possibility that they could exist in intergalactic space, in which case some of the above arguments can be circumvented. There is no {\it current} evidence for such objects but their existence only entails a small extrapolation from the available data, so it seems surprising that this possibility has been neglected in the literature. The purpose of this paper is to consider some of their observational consequences: in particular, their dynamical and lensing effects, the influence of their accretion-generated luminosity on the thermal history of the Universe, and the annihilation of the WIMPs expected to form a halo around them.

Any intergalactic SLABs must have formed independently of galaxies and would probably need to be {\it primordial} in origin. It has long been argued that primordial black holes (PBHs) could have formed in the early radiation-dominated Universe~\cite{1967SvA....10..602Z, Hawking:1971ei, Carr:1974nx}. They are not usually expected to be as large as SLABs but in principle their mass could be anything up to the horizon mass at the time of matter-radiation equality, which is of the order of $10^{17}\,M_{\odot}$. One fairly generic scenario in which the PBH mass function extends up to the SLAB range has been suggested by Vilenkin and colleagues~\cite{Garriga:2015fdk, Deng:2016vzb, Deng:2017uwc, Deng:2020mds, Deng:2018cxb, Deng:2020pxo}.

In this context, it is important to stress that even the SMBHs in galactic nuclei could conceivably have been seeded by PBHs. The conventional assumption is that the SMBHs in galactic nuclei form as a result of dynamical processes {\it after} the galaxies themselves~\cite{Rees:1984si}. Specific scenarios are discussed in Sec.~\ref{sec:Formation-of-SLABs}. However, observations of quasars at redshifts $z \gtrsim 6$ suggest that SMBHs larger than $10^{9}\,M_{\odot}$ were already present when the Universe was less than a billion years old~\cite{2020arXiv200613452Y}. Generating BHs this large so early, while certainly possible~\cite{Begelman:2006db}, is challenging. Therefore it is interesting to consider the possibility that the SMBHs in galaxies formed {\it before} galaxies, in which case they could be primordial. However, the PBHs would inevitably have grown enormously through accretion since matter-radiation equality, which is why they should only be regarded as seeding the SMBHs.

If SLABs are of primordial origin, this raises an interesting link with the suggestion that PBHs could provide the dark matter (see Ref.~\cite{Carr:2020xqk} for a recent review). Although SLABs themselves clearly cannot do this, since they are too large to reside in galactic halos, it is possible that PBHs provide the dark matter in a much lower mass range. This is because PBHs formed in the radiation-dominated era and are therefore non-baryonic, circumventing the usual bound on the density of baryonic matter. For example, PBHs could have formed from quantum fluctuations which were produced and re-entered the horizon during inflation~\cite{Carr:1993aq, Ivanov:1994pa, GarciaBellido:1996qt}. Although the SLAB and DM populations might be distinct, they could be related if PBHs have an extended mass function.

While there is no definitive evidence that PBHs provide the dark matter, there is a huge literature discussing constraints on their contribution to the dark density~\cite{Carr:2016drx, Carr:2020gox}, these being associated with a wide variety of effects: quantum evaporation, gravitational lensing, dynamical effects, accretion and influence on cosmic structures. These studies show that there are only four mass windows in which PBHs could have an ``appreciable'' density: 
	(A) the asteroid mass range with $10^{16} < M / \grm < 10^{18}$; 
	(B) the lunar mass range with $10^{20} < M / \grm < 10^{24}$; and
	(C) the intermediate mass range with $10 < M /\,M_{\odot} < 10^{2}$; 
	(D) the stupendous mass range with $10^{12} < M /\,M_{\odot} < 10^{18}$.
While the first three windows have been well studied, the last window{\;---\;}which corresponds to the SLAB range{\;---\;}has been almost completely neglected.

The apparent lack of constraints on SLABs probably just reflects the fact that very little attention has been paid to their possible existence, except perhaps in galactic nuclei. Nevertheless, they could have striking observational consequences and the purpose of this paper is to examine some of these. Even though they could not explain the dark matter, they could still represent a smoothly distributed dark intergalactic contribution, so it is interesting to know how large this could be.

If SLABs are pregalactic, a particularly interesting constraint would be associated with their accretion of pregalactic gas. This problem was originally studied in Ref.~\cite{10.1093/mnras/194.3.639}, on the assumption that the BHs accrete at the Bondi rate. However, this neglected the fact that steady-state Bondi is inappropriate for very large BHs because the Bondi timescale exceeds the cosmic expansion time. It is still not clear how to correct for this but we will attempt to address this issue. Subsequently, many other authors~\cite{Ricotti:2007au, Ricotti:2007jk, Ali-Haimoud:2016mbv} have studied PBH accretion but their analyses only apply for masses below around $10^{4}\,M_{\odot}$, in part because of the failure of the steady-state assumption. However, it is clear that the accretion constraints do not suddenly cut off above this mass.

If the PBH dark-matter fraction is low, some other candidate must dominate and the most studied is a weakly interacting massive particle (WIMP). Each PBH would then provide a seed around which a halo of WIMPs would form, the gamma-rays from their annihilations then implying stringent bounds on combined dark-matter scenarios~\cite{Eroshenko:2016yve, Boucenna:2017ghj, Eroshenko:2019pxt, Adamek:2019gns,Cai:2020fnq}. A dark matter halo around BH binaries would also alter the expected gravitational-wave signal~\cite{Edwards:2019tzf, Kavanagh:2020cfn}. Previous work in this context has focussed on BHs in the mass ranges below $10^{3}\,M_{\odot}$ but in this paper we will extend these arguments to the SLAB range. In a separate paper~\cite{Visinelli:2020B}, we discuss this problem in a more general context, covering the entire mass range $10^{-12}$ -- $10^{12}\,M_{\odot}$ and distinguishing between the Galactic and extragalactic limits. Here we focus on the SLAB range, where only the extragalactic limit is relevant.

Regardless of their origin, SLABs could also play an important r{\^o}le in the presence of light bosonic fields. Such bosons are expected to form condensates around the holes and their quantum fluctuations would populate the quantum levels of the boson clouds~\cite{Arvanitaki:2009fg}. If the BHs spin, a fraction of the rotational energy might be transferred to the surrounding boson cloud through superradiance~\cite{Compton:1923zz, Frank:1937fk}, even if the bosons only provide a small fraction of the dark matter. This superradiant instability of rotating BHs has attracted increasing interest as a possible probe of light bosons~\cite{Arvanitaki:2010sy}. However, this requires the BHs to spin appreciably, which is not expected for PBHs~\cite{Mirbabayi:2019uph, DeLuca:2019buf, DeLuca:2020bjf} except in non-standard cosmological scenarios~\cite{Harada:2017fjm, Carr:2018nkm}. The effect of a boson cloud on the angular size of the shadow of M87$^{*}$ has also been considered in Ref.~\cite{Davoudiasl:2019nlo}, a lower bound on the mass of the light boson being placed once the spin and the size of the BH shadow is measured with sufficient accuracy.

The plan of this paper is as follows. In Sec.~\ref{sec:Formation-of-SLABs} we discuss how SLABs might form either primordially or in galactic nuclei and clarify the relationship between these two scenarios. In Sec.~\ref{sec:Lensing-Constraints-on-SLABs} we review the lensing bounds on SLABs and in Sec.~\ref{sec:Dynamical-Constraints-on-SLABs} we review the dynamical bounds. Most of these bounds have been reported before but we make them more precise and correct some errors in previous treatments. In Sec.~\ref{sec:Accretion-Constraints-on-SLABs} we derive accretion constraints, developing some much earlier work but stressing that our approach faces some difficulties. In Sec.~\ref{sec:SLAB-Constraints-from-WIMP-Annihilations} we consider the constraints associated with the annihilation of WIMPs expected to form halos around SLABs, extending previous studies for a much lower PBH mass range. In Sec.~\ref{sec:SLABs-and-Light-Bosons} we focus on constraints on the mass of light bosons if the SLABs are rotating. Some general conclusions are drawn in Sec.~\ref{sec:Results-and-Discussion}.

\section{Formation of SLABs}
\label{sec:Formation-of-SLABs}

Although we should stress at the outset that there is no current evidence for SLABs, the emphasis of this paper being on constraints, we will now discuss their possible formation scenarios. There are two classes of model. The first{\;---\;}perhaps most natural{\;---\;}possibility is that they form in galactic nuclei, although we will see that there are both theoretical and observational reasons for believing that the mass function of such SMBHs does not extend beyond $10^{11}\,M_{\odot}$. Indeed, this is one reason for taking this to be the lower mass limit in our definition of a SLAB. The second{\;---\;}more speculative{\;---\;}possibility is that they form primordially (\ie~within the first few minutes of the Universe). One does not usually envisage PBHs being in the SLAB range and there are arguments that their masses should not extend beyond $\sim 10^{5}\,M_{\odot}$. However, there are some scenarios for producing PBHs larger than this and, as we will see, very massive ones will inevitably grow enormously through accretion, so one must distinguish between the initial and final PBH mass. It is also possible that that the SMBHs in galactic nuclei were themselves seeded by PBHs, so the distinction between the two classes of explanation is not clear-cut. In principle, SMBHs could be located in intergalactic space rather than galactic nuclei but one then needs to explain why such enormous objects do not act as seeds for galaxies or clusters anyway.

\subsection{Postgalactic SLABs in galactic nuclei}

If the SMBHs in galactic nuclei are generated by dynamical processes {\it after} the formation of the galaxies, there are two possible pathways: direct collapse to SMBHs~\cite{Habouzit:2016nyf} or super-Eddington growth~\cite{Pezzulli:2016pxl}. Both of these have been explored in a series of papers by Agarwal {\it et al.}~\cite{Agarwal:2012qm, Agarwal:2013xfa, Agarwal:2014hla, Johnson:2014jja} but they are not without difficulties. In the former case, the seeds are rare; in the latter case, they should be ultra-luminous and visible in deep X-ray surveys. Other scenarios include the direct collapse of gas clouds in the center of minihalos~\cite{Loeb:1994wv}, the rapid accretion experienced by a black hole moving inside a star cluster~\cite{Alexander:2014noa}, and the formation out of supermassive stars~\cite{Hosokawa:2013mba, Reisswig:2013sqa} and dark stars~\cite{Freese:2015mta}.

Whatever the scenario, consistency between X-ray and optical data at high $z$ and the observed SMBH mass function at low $z$ implies an upper limit on the mass. This limit was first studied by Natarajan and Treister~\cite{Natarajan:2008ks}, who pointed out that consistency requires the mass function to steepen above around $10^{9}\,M_{\odot}$, the UMBH scale, this being explained by a self-regulation mechanism that limits the mass. One expects a link between star formation and black hole fuelling and they discuss several mechanisms which explain the observed $M$--$\.\sigma^{4}$ relation. The clearing out the nuclear region by the growing black hole leads to an upper limit of $\sim 10^{10}\,M_{\odot}$ and the outflow from the accretion may also halt the flow, the mass limit in this case being similar but sensitive to the halo spin and disc mass fraction. Extrapolating the $M$--$\.\sigma^{4}$ relation to high $\sigma$ suggests that UBMHs might be hosted by massive high-luminosity galaxies with large velocity dispersions in the centres of clusters. This is also a natural implication of the accretion and merger of smaller SMBHs. The highest velocity dispersion currently observed is $\sim 400$\,km/s and the limit would then be $\sim 10^{10}\,M_{\odot}$. However, for a cD (central Dominant) galaxy with a velocity dispersion of $\sim 700$\,km/s, it would be $\sim 10^{11}\,M_{\odot}$.
 
The model of King~\cite{10.1093/mnrasl/slv186} assumes that the SMBH is surrounded by an accretion disc which is self-gravitating outside some radius $R_{\rm sg}$. Gas in the outer region then cools fast enough for self-gravity to lead to star formation. Most of the gas initially outside $R_{\rm sg}$ either goes into these stars or is expelled by them on a near-dynamical timescale. This implies that the outer radius of the accretion disc cannot exceed $R_{\rm sg}$. However, the inner radius must be at least as large as the ISCO (innermost stable circular orbit) and this precludes disc formation for SMBHs larger than some limit which depends on the ratio $L/L_{\rm ED}$ and the value of $\alpha$. The limit is $5 \times 10^{10}\,M_{\odot}$ for $L = L_{\rm ED}$ and $\alpha = 0.1$. This applies for any SMBHs in quasars or AGN, since these must have accretion discs. For maximal prograde spin, it becomes $2.7 \times 10^{11}\,M_{\odot}$, which is the absolute maximum for an accreting SMBH. At masses just below the limit, the disc luminosity of a field galaxy is likely to be below the Eddington limit and not strong enough to trigger the feedback underlying the $M$--$\.\sigma^{4}$ relation, so SMBHs close to the limit can evolve above this.

The model of Yazdi and Niayesh~\cite{Yazdi:2016fyt} modifies the Shakura-Sunyaev model for radiation-pressure-dominated disks. They calculate the radius at which $P_{\rm rad} = P_{\rm gas}$ as a function of $\dot{M} / \dot{M}_{\rm ED}$ for different values of $M$ and the $\alpha$ parameter. The more efficient cooling then keeps the disk in equipartition with $P_{\rm rad} = P_{\rm gas}$ beyond this radius. As in the King model, self-gravity becomes important in this region, so the matter clumps there and no longer accretes onto the SMBH. The larger the mass, the closer $R_{\rm sg}$ gets to the ISCO radius and there will no longer be accretion when they become equal. This sets a redshift-dependent upper limit to the SMBH mass. This lies in the range $( 2$ -- $5 ) \times 10^{11}\,M_{\odot}$ at $z = 0.1$ and $( 0.5$ -- $1.5) \times 10^{11}\,M_{\odot}$ at $z = 10$ for non-rotating BHs but it is reduced for highly rotating BHs. Both their argument and King's are based on the Toomre instability of $\alpha$-disks~\cite{Toomre:1964zx}.

\subsection{Primordial SLABs in galactic nuclei}

Generating a SMBH with a mass of $10^{10}\,M_{\odot}$ by a redshift of $z = 7$ (as observed) is challenging. It is therefore interesting to consider the possibility that the SMBHs formed {\it before} galaxies~\cite{Bean:2002kx, Duechting:2004dk, Clesse:2015wea, Khlopov:2004tn} and this leads to three possible scenarios.

(i) The first possibility is that the PBHs underwent very little accretion and were themselves supermassive, so that they can be directly identified with the SMBHs. In this case, they could also help to {\it generate} galaxies through either the seed or Poisson effect, the fluctuations growing by a factor $\sim 10^{3}$ between the time of matter-radiation equality and galaxy formation~\cite{Carr:2018rid}. This naturally explains the proportionality between the SMBH and galaxy mass and also provides an early mode of galaxy formation that might be important for the reionisation of the Universe~\cite{Chevallard:2014sxa}. In this case, the galaxy mass function directly reflects the PBH mass function and the latter might conceivably extend to the SLAB scale.

(ii) The second possibility is that the PBHs had a more modest mass and then grew through Eddington-limited accretion. This scenario was first suggested by Bean and Magueijo~\cite{Bean:2002kx}, although they overestimated the amount of accretion in the early radiation-dominated phase. They argued that it needs a very narrow PBH mass function to reproduce the observed distribution of SMBHs and Kawasaki {\it et al.}~\cite{Kawasaki:2012kn} suggested a specific inflationary scenario to account for this. However, most of the accretion still occurs after decoupling, so it may be difficult to distinguish this observationally from a scenario in which the original BHs are non-primordial. In both cases, one would expect a lot of radiation to have been generated and this would contribute to the observed X-ray background~\cite{Soltan:1982vf}. In a recent work, Nunes and Pacucci~\cite{Nunes:2020yij} have analysed the most massive high-redshift SMBHs and argued that the seed mass would need to be at least $10^{4}\,M_{\odot}$ to explain the observations. They find that $M > 10^{12}\,M_{\odot}$ SLABs are theoretically possible but that feedback and galaxy dynamics effects will have important observational consequences above $10^{13}\,M_{\odot}$. It is unclear whether an object of $10^{17}\,M_{\odot}$ would be observationally compatible with other SMBH data. For example, its gravitational-wave signal would be out of the range even of LISA and its electromagnetic signal would be uncertain.

(iii) The third possibility is that the PBHs had an even more modest mass and generated the SMBHs in galactic nuclei through the seed or Poisson effect. For example, to produce a SMBH of mass $10^{8}\,M_{\odot}$ by $z \sim 4$, one requires $M \sim 10^{5}\,M_{\odot}$ for the seed effect or $M \sim 10^{2}\,M_{\odot}$ for the Poisson effect if the PBHs provide the dark matter. To produce a SLAB of $10^{10}\,M_{\odot}$, one would require $M \sim 10^{7}\,M_{\odot}$ for the seed effect or $M \sim 10^{4}\,M_{\odot}$ for the Poisson effect. However, this only produces a gravitationally bound region and one still has to explain how a region containing a central large PBH or a cluster of intermediate mass PBHs can evolve to a single SMBH. Accretion and merging would clearly be important and only some fraction of the bound region might end up in the central BH.

\subsection{Intergalactic primordial SLABs}

Finally we consider the possibility that there is an intergalactic population of SLABs, not necessarily related to the SMBHs in galactic nuclei. In this case, they would almost certainly need to be primordial. One scenario in which primordial SLABs form naturally has been suggested by Vilenkin and colleagues~\cite{Garriga:2015fdk, Deng:2016vzb,Deng:2017uwc, Deng:2020mds, Deng:2018cxb, Deng:2020pxo}. In this case, the PBHs are formed by bubbles nucleated during inflation and their mass function scales as $f( M ) \propto M^{-1/2}$, so the value of $f( M )$ at $10^{11}\,M_{\odot}$ is uniquely determined by the total PBH density. For example, if PBHs of $10\,M_{\odot}$ provide the dark matter, one would have $f \sim 10^{-5}$ in the SLAB range. This mass function would also apply if the PBHs formed from the collapse of cosmic strings, although existing constraints on the string parameter would exclude them providing the dark matter. The power spectrum invoked in the ``thermal history'' model of Ref.~\cite{Carr:2019kxo} would also produce a significant SLAB density, although the spectrum needs to be cut off at some mass-scale to be consistent with large-scale structure observations.

Whatever the source of very massive PBHs, they would inevitably increase their mass by accretion of pregalactic gas [cf. case B(ii) above]. A simple Bondi accretion analysis suggests that such holes go through an Eddington accretion phase which ends at some time $t_{\rm ED}$ after the Big Bang. During this phase, each BH doubles its mass on the Salpeter timescale, $t_{\Srm} \approx 4 \times 10^{8}\,\epsilon\,$yr where $\epsilon$ is the luminosity efficiency~\cite{Salpeter:1964kb}. This corresponds to the age of the Universe at a redshift $z_{\Srm} \approx 40\.( \epsilon / 0.1 )^{2/3}$. Therefore, for $\epsilon = 0.1$ one expects the mass of the BH to increase by a factor $\exp( t_{\rm ED} / t_{\Srm} ) \approx \exp[ ( 40 / z_{\rm ED} )^{3/2} ]$, which is very large for $z_{\rm ED} \ll 40$. For example, the growth up to $z = 7$ would be of order $10^{6}$. In Sec.~\ref{sec:Accretion-Constraints-on-SLABs}, we will calculate the value of $z_{\rm ED}$ as a function of the PBH mass $M$ and density parameter $\Omega_{\rm PBH}( M )$.

Understanding accretion is crucial in discussing the origin of SLABs, since it is unlikely that PBHs could be in the SLAB mass range {\it initially}. This is because the formation of PBHs from primordial inhomogeneities requires that the power spectrum of the curvature fluctuations be $\Ocal( 1 )$, whereas observational constraints from the observed CMB anisotropies and spectral distortions excludes this above a mass of $M_{\rm max} \sim 10^{10}\,M_{\odot}$. One therefore needs a growth factor of at least $M_{\rm SLAB} / M_{\rm max}$. Since the initial mass may only be tiny fraction of the final mass, and since most of the mass increase occurs at the end of the Eddington phase (\ie~long after matter-radiation equality), it is not so crucial whether the initial seed was primordial. There is also the semantic issue of how late a black hole can form and still be described as {\it primordial}. One usually assumes that PBHs form in the radiation-dominated era but the crucial feature is that they form from $\Ocal( 1 )$ density fluctuations rather than via the fragmentation of much larger regions (like galaxies) during the matter-dominated era.\\

\section{Lensing and CMB Constraints on SLABs}
\label{sec:Lensing-Constraints-on-SLABs}

If one has a population of compact objects with mass $M$ and density $\Omega_{\Crm}$, then the probability $P$ of one of them image-doubling a source at redshift $z \approx 1$ and the separation $\theta_{\srm}$ between the images are~\cite{1973ApJ...185..397P}
\begin{equation}
	P
		\approx
					(
						0.1
						-
						0.2
					)\,
					\Omega_{\Crm}
					\, ,
		\;\;
	\theta_{\srm}
		\approx
					5 \times 10^{-6}\,
					(
						M / M_{\odot}
					)^{1/2}\,{\rm arcsec}
					\, .
\end{equation}
One can therefore use the upper limit on the frequency of macrolensing for different image separations to constrain $\Omega_{\Crm}$ as a function of $M$. The usual approach is to derive a ``detection volume'', defined as the volume between the source and the observer within which the lens would need to lie to produce an observable effect. Limits are then obtained by adding the detection volume for each source and comparing this to the volume per source expected for a given $\Omega_{\Crm}$.

There have been several optical and radio surveys to search for multiply-imaged quasars~\cite{Bahcall:1991qs}. In particular, Hewitt~\cite{10.1007/978-94-009-0295-4_43} used VLA observations to infer $\Omega_{\Crm}\.( 10^{11}$ -- $10^{14}\,M_{\odot} ) < 0.4$, Nemiroff~\cite{PhysRevLett.66.538} used optical QSO data to infer $\Omega_{\Crm}( M > 10^{9.9}\,M_{\odot} ) < 1$ and $\Omega_{\Crm}( M > 10^{10.3}\,M_{\odot} ) < 0.4$, and Surdej {\it et al.}~\cite{1993AJ....105.2064S} used data on 469 highly luminous quasars to infer $\Omega_{\Crm}( 10^{10}$ -- $10^{12}\,M_{\odot} ) < 0.02$. Most of the recent emphasis has been on pushing these macrolensing limits down to lower masses but in the present context we are interested in larger masses~\cite{Virbhadra:1999nm, Virbhadra:2008ws}.

SLABs would generate ``cluster-like'' anisotropies in the CMB sky, with a lensing signal scaling proportional to the mass. Any object with $M > 10^{16}\,M_{\odot}$ and located near the peak of the CMB lensing kernel (\ie~at redshifts $z = 0.5$ -- $10$) should show up as a huge signal, possibly visible by eye in the SPT or ACT lensing maps. In the $10^{14}$ -- $10^{15}\,M_{\odot}$ range, the issue is less clear, since usually one only obtains good signal-to-noise by stacking the clusters (Horowitz, private communication).

The Vilenkin {\it et al.} scenario, in which the PBHs form from inflationary bubbles during, is particularly interesting in this context. There are no large density fluctuations outside the bubbles, so spectral distortions are localised in small angular patches on the sky. They discuss observational constraints on the model due to spectral distortions and their calculations extend up to $10^{20}\,M_{\odot}$. They find that the largest BH that one can expect to find in the observable region has initial mass $\sim 10^{14}\,M_{\odot}$~\cite{Deng:2018cxb,
Deng:2020pxo}.

\section{Dynamical Constraints on SLABs}
\label{sec:Dynamical-Constraints-on-SLABs}

The most clear-cut constraints on the PBH fraction $f_{\rm PBH}$ in the mass range above $10^{5}\,M_{\odot}$ are dynamical and summarised in Fig.~\ref{fig:SLAB-Constraints}. All of them have been discussed before, although we will need to refine some of the arguments for the SLAB range. Before reviewing the arguments, we point out an obvious {\it lower} limit on $f_{\rm PBH}( M )$, which has been termed the ``incredulity limit''~\cite{Carr:1997cn}. The PBH scenario is only interesting if there is at least one PBH in the relevant environment, be it a galactic halo or a cluster of galaxies or the entire observable Universe. This corresponds to the condition 
\begin{equation}
	\label{eq:IL}
	f_{\rm PBH}( M )
		\geq
					\frac{ M }{ M_{\Erm} }
					\, ,				
\end{equation}
where $M_{\Erm}$ is the mass of the environment, taken to be $10^{12}\,M_{\odot}$ for a galactic halo, $10^{14}\,M_{\odot}$ for a cluster of galaxies and $10^{22}\,M_{\odot}$ for the observable Universe (\ie~the mass within the current particle horizon). This specifies the right edge of some constraints shown in Fig.~\ref{fig:SLAB-Constraints}. However, the scenario is not necessarily excluded beyond this edge but merely irrelevant to the associated dark matter problem. For example, it makes no sense to postulate the halo being made of objects as big as the halo itself, so SLABs are clearly outside the galactic incredulity limit, though not necessarily the cluster one.

\subsection{Dynamical friction limit}

Halo BHs with $M > 10^{5}\,M_{\odot}$ would be dragged into the nucleus of the Galaxy by the dynamical friction of various stellar populations. These holes can then merge to form a single SMBH larger than the observed mass of $4 \times 10^{6}\,M_{\odot}$ unless $f_{\rm PBH}( M )$ is suitably constrained~\cite{Carr:1997cn}. We include this limit in Fig.~\ref{fig:SLAB-Constraints} but do not give an explicit expression for it since it is complicated. This is because there are different sources of friction and the limit depends on parameters such as the halo core radius. Also it may be weakened if black holes can be ejected from the Galactic nucleus by 3-body effects~\cite{Xu:1994vb}. The limit bottoms out at $M \sim 10^{7}\,M_{\odot}$ with a value $f_{\rm PBH} \sim 10^{-4}$.

\subsection{Destruction of galaxies in clusters}

If the dark matter in clusters comprises SLABs, then galaxies would be heated up and eventually destroyed by encounters with them. The associated limit can be expressed as~\cite{Carr:1997cn}
\begin{equation}
	f_{\rm PBH}( M )
		<
					\!\begin{cases}
						(
							M / 10^{10}\,M_{\odot}
						)^{-1}
							& \!\!(10^{10} < M / M_{\odot} < 10^{11} )
						\, ,\\[1mm]
						0.1
							& \!\!( 10^{11} < M / M_{\odot} < 10^{13} )
						\, ,\\[1mm]
						(
							M / 10^{14}\,M_{\odot}
						)
							& \!\!( 10^{13} < M / M_{\odot} < 10^{14} )
						\, ,
					\end{cases}_{_{_{_{}}}}
\end{equation}
although it depends sensitively on the mass and the radius of the cluster. The last expression corresponds to the incredulity limit of one SLAB per cluster. Below this line, the SLABs are not confined to clusters and the tidal distortion limit is weakened by factor of $100$, corresponding to the ratio of the cluster density to the cosmological background density, and is uninteresting.

There is also a constraint from the lack of unexplained tidal distortions of galaxies in clusters. If the fraction of distorted galaxies at any time is $\Delta_{\grm}$, this limit takes the form~\cite{Carr:1997cn}
\begin{equation}
	f_{\rm PBH}( M )
		<
					\lambda^{-3}\.\Delta_{\grm}
					\left(
						\frac{ \rho_{\grm} }{ \rho_{\rm DM} }
					\right)
		\approx
					20
					\left(
						\frac{ \lambda }{ 3 }
					\right)^{\!-3}
					\Delta_{\grm}
\end{equation}
where $\lambda \approx 3$ represents the difference between distortion and disruption. So one has an interesting limit providing the fraction of unexplained distorted galaxies is less than $5$\%. This limit only applies for $M > \lambda^{-3} M_{\grm} \approx 3 \times 10^{9}\,M_{\odot}$ since otherwise the tidal distortion radius is smaller than $R_{\grm}$. In particular, Van den Bergh~\cite{1969Natur.224..891V} concluded from observations of the Virgo cluster, which has a mass $10^{15}\,M_{\odot}$ and only 4 of 73 cluster members with unexplained tidal distortions, that the dark matter could not be in the form of compact object in the mass range $( 10^{9}$ -- $10^{13} )\,M_{\odot}$.This assumes that the distortion only persists during the encounter itself, the tidal stretching being reversed as the SLAB recedes.

\subsection{Large-scale structure limits}

PBHs larger than $10^{2}\,M_{\odot}$ cannot provide dark matter but Carr and Silk~\cite{Carr:2018rid} have studied how such PBHs could generate cosmic structures through the `seed' or `Poisson' effect even if $f_{\rm PBH}$ is small. If a region of mass $\tilde{M}$ contains PBHs of mass $M$, the initial fluctuation is
\begin{equation}
	\label{eq:initial}
	\delta_{i}
		\approx
					\begin{cases}
						M / \tilde{M}
							& ( \text{seed} )
						\, ,\\[1.5mm]
						(
							f_{\rm PBH}\,M / \tilde{M}
						)^{1/2}
							& ( \text{Poisson} )
						\, .
					\end{cases}
\end{equation}
If $f_{\rm PBH} = 1$, the Poisson effect dominates for all $\tilde{M}$\,; if $f \ll 1$, the seed effect dominates for $\tilde{M} < M / f_{\rm PBH}$\, . In either case, the fluctuation grows as $z^{-1}$ from the redshift of DM domination ($z_{\rm eq} \approx 4000$), so the mass binding at redshift $z_{\Brm}$ is
\begin{equation}
	\label{eq:bind}
	\tilde{M}
		\approx
					\begin{cases}
						4000\,
						M\.
						z_{\Brm}^{-1}
							& ( \text{seed} )
						\, ,\\[1.5mm]
						10^{7}\,
						f_{\rm PBH}\,
						M\.
						z_{\Brm}^{-2}
							& ( \text{Poisson} )
						\, .
					\end{cases}
\end{equation}
This assumes the PBHs have a monochromatic mass function. If it is extended, the situation is more complicated because the mass of the effective seed for a region may depend on the mass of that region~\cite{Carr:2018rid}.

\begin{figure}[t]
	\includegraphics[width = 1.05\linewidth]{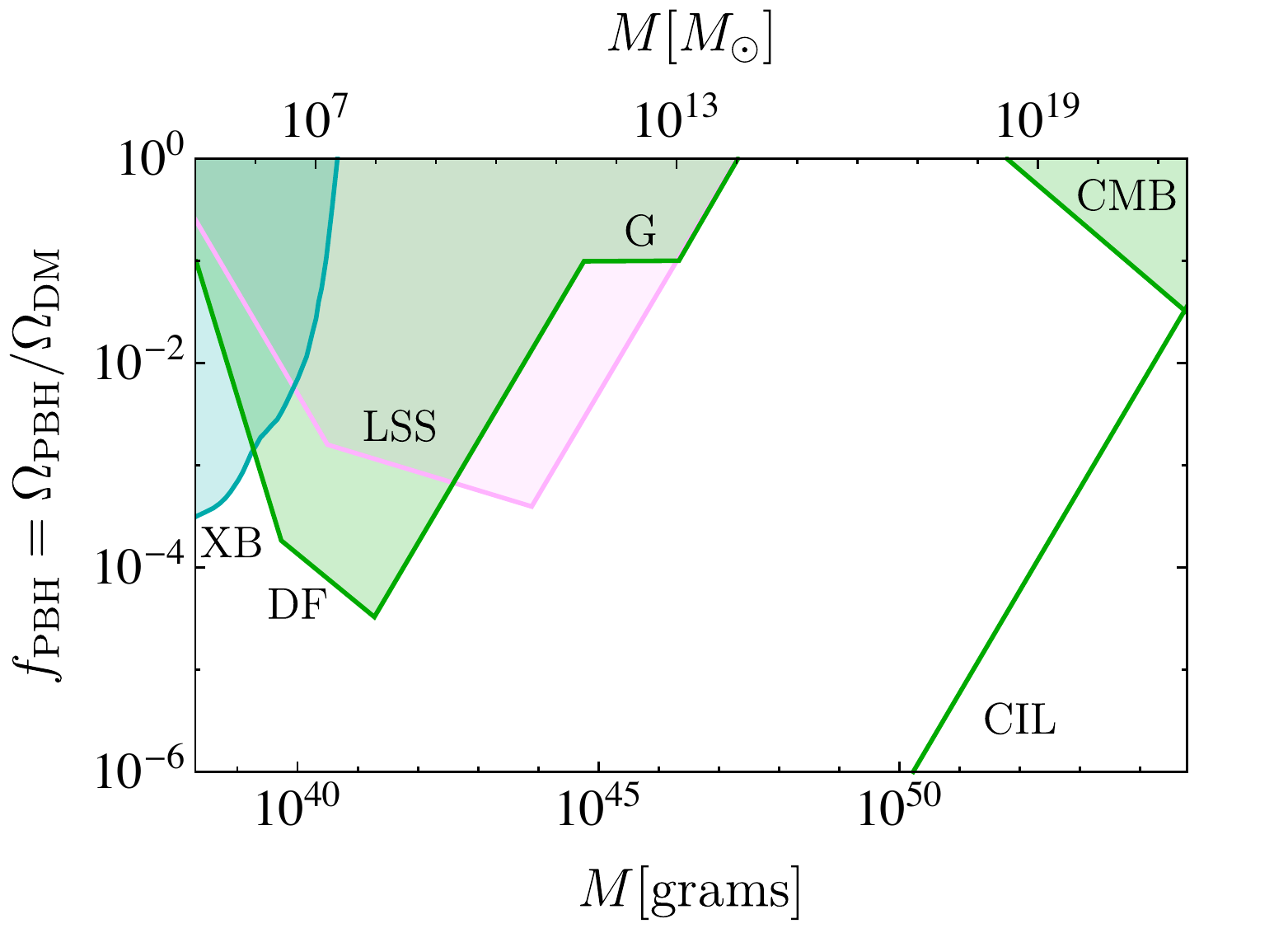} 
	\caption{SLAB constraints on $f_{\rm PBH}( M )$ 
			for a monochromatic mass function. Dynamical limits derive from 
			halo dynamical friction (DF), 
			galaxy tidal distortions (G)
			and the CMB dipole (CMB).
			Large-scale structure limits derive from requiring that 
			various cosmological structures do not form earlier than observed (LSS).
			Accretion limits come from X-ray binaries (XB).
			The cosmological incredulity limit (CIL) corresponds to 
			one PBH within the cosmological horizon.
			Based in part on Ref.~\cite{Carr:2020xqk}.}
	\label{fig:SLAB-Constraints}
\end{figure}

Even if PBHs do not play a r{\^o}le in generating cosmic structures, one can still place interesting upper limits on the fraction of dark matter in them by requiring that various types of structure do not form too early~\cite{Carr:2018rid}. For example, if we apply the above argument to Milky-Way-type galaxies, assuming these must not bind before a redshift $z_{\Brm} \sim 3$, we obtain
\begin{equation}
	\label{eq:galaxy}
	f_{\rm PBH}( M )
		<
					\!\begin{cases}
						( M / 10^{6}\,M_{\odot} )^{-1}
							& \!\!\!\!\!( 10^{6} < M / M_{\odot} \lesssim 10^{9} )
						\, ,\\[1.5mm]
						( M / 10^{12}\,M_{\odot} )
							& \!\!\!\!\!( 10^{9} \lesssim M / M_{\odot} < 10^{12} )
						\. .
					\end{cases}_{_{_{_{}}}}
\end{equation}
This limit bottoms out at $M \sim 10^{9}\,M_{\odot}$ with a value $f_{\rm PBH} \sim 10^{-3}$. The first expression can be obtained by putting $\tilde{M} \sim 10^{12}\,M_{\odot}$ and $z_{\Brm} \sim 3$ in Eq.~\eqref{eq:bind}. The second expression corresponds to having just one PBH per galaxy and is also the line above which the seed effect dominates the Poisson effect ($f_{\rm PBH} < M / \tilde{M}$). (This limit also arises because competition from other seeds implies that the mass bound is at most $f_{\rm PBH}^{-1}\,M$.) Similar constraints are associated with the first bound clouds, dwarf galaxies and clusters of galaxies, all the limits being shown by the LSS line in Fig.~\ref{fig:SLAB-Constraints}.

There is also a Poisson constraint associated with observations of the Lyman-$\alpha$ forest~\cite{Afshordi:2003zb, Murgia:2019duy}, which has a similar form to Eq.~\eqref{eq:galaxy} but with the lower limit of $M \sim 10^{6}\,M_{\odot}$ reduced to $M \sim 10^{2}\,M_{\odot}$. The constraint reaches a minimum value $f \sim 10^{-5}$ at $10^{7}\,M_{\odot}$. However, it is unclear how to extend this limit to larger values of $M$. The galactic incredulity limit need not apply because Lyman-$\alpha$ forest observations would be affected even if there was less than one PBH per halo. If the limit were flat beyond $10^{7}\,M_{\odot}$, it would be strongest constraint in the SLAB region of Fig.~\ref{fig:SLAB-Constraints}.
 However, the relationship between the Poisson and seed effects is rather subtle in this context. The latter can still be important below the incredulity limit but only influences a small fraction of regions, so the interpretation of the observations is unclear.

\subsection{CMB dipole limit}

If there were a population of huge intergalactic PBHs with density parameter $\Omega_{\rm IG}( M )$, which is the likely situation with SLABs, each galaxy would have a peculiar velocity due to its gravitational interaction with the nearest one~\cite{1978ComAp...7..161C}. If the objects were smoothly distributed, the typical distance between them would be~\cite{Carr:2020gox}
\begin{equation}
	\label{eq:distance}
	d
		\approx
					40\;\Omega_{\rm IG}( M )^{-1/3}
					\left(
						\frac{ M }{ 10^{16}\,M_{\odot} }
					\right)^{\!1/3}\!
					\left(
						\frac{ h }{ 0.7 }
					\right)^{\!\!-2/3}
					{\rm Mpc}
					\, ,
\end{equation}
where $h$ is the Hubble constant in units of $100\,{\rm km}\,\mathrm s^{-1}\,\mathrm {Mpc}^{-1}$, and this should also be the expected distance of the nearest one. Over the age of the Universe $t_{0}$, this should induce a peculiar velocity in the Milky Way of $V_{\rm pec} \approx G M\,f( \Omega_{\mrm} )\,t_{0} / d^{2}$ where $\Omega_{\mrm} \approx 0.3$ is the total matter density parameter (dark plus baryonic) and $f( \Omega_{\mrm} ) \approx \Omega_{\mrm}^{0.6}$. Since the CMB dipole anisotropy shows that the peculiar velocity of the Milky Way is only $400\,{\rm km}\,\srm^{-1}$, one infers
\begin{equation}
	\label{eq:dipole}
	\Omega_{\rm IG}
		<
					\left(
						\frac{ M }{ 2 \times 10^{16}\,M_{\odot} }
					\right)^{\!-1/2}\!
					\left(
						\frac{ t_{0} }{ 14 \,{\rm Gyr} }
					\right)^{\!\!-3/2}\!
					\left(
						\frac{ h }{ 0.7 }
					\right)^{\!-2}
					\, .
\end{equation}
This scenario is interesting only if there is at least one such object within the observable Universe and this corresponds to the lower limit
\begin{equation}
	\label{eq:CIL}
	\Omega_{\rm IG}( M )
		>
					2 \times 10^{-8}
					\left(
						\frac{ M }{ 10^{16}\,M_{\odot} }
					\right)\!
					\left(
						\frac{ t_{0} }{ 14 \,{\rm Gyr} }
					\right)^{\!-3}
					\left( \frac{h}{0.7} \right)^{-2}
					\, ,
\end{equation}
where we have taken the horizon scale to be $d \approx 3\,c\,t_{0} \approx 10\,h^{-1}\,{\rm Gpc}$. This intersects Eq.~\eqref{eq:dipole} at
\begin{equation}
	M
		\approx 
					1.1 \times 10^{21}
					\left(
						\frac{ t_{0} }{ 14 \,{\rm Gyr} }
					\right)
					M_{\odot}
					\, ,
\end{equation}
so this corresponds to the largest possible BH within the visible Universe.

\section{Accretion Constraints on SLABs}
\label{sec:Accretion-Constraints-on-SLABs}

Clearly there would be interesting constraints on SLABs due to their accretion at the present or recent epochs but the form of these constraints depends on their environment and is complicated. For example, if the SLABs fall within larger-scale cosmic structures, this will have important consequences for their velocity and spatial distribution. If they reside in galactic nuclei, or even seed galaxies, they would accrete local gas and stars in a manner which has been studied in much previous literature. If they reside outside galaxies, or even outside clusters of galaxies, they would still accrete intergalactic gas, the consequences of which depend on the (somewhat uncertain) state of the intergalactic medium. Here we study the accretion of gas by SLABs before the formation of galaxies or other cosmic structures. They may also accrete dark matter in this period, this being relevant to the considerations of Sec.~\ref{sec:SLAB-Constraints-from-WIMP-Annihilations}, but we neglect that process here.

\subsection{Pregalactic gas accretion}

Our analysis is based on the Bondi accretion formula and covers a wide range of epochs, although only PBHs could exist all the way back to matter-radiation equality at $t_{\rm eq} = 2.4 \times 10^{12}\,$s or $\zeq \approx 3400$. Before $t_{\rm eq}$, the sound-speed is $c_{\srm} = c / \sqrt{3}$, where $c$ is the speed of light, and one can show that there is very little accretion~\cite{Carr:1974nx}. After $t_{\rm eq}$, the accretion radius is increased (since $c_{\srm}$ falls below $c$) and the accretion rate is larger. However, the problem is complicated because the BH luminosity will boost the matter temperature of the background Universe above the standard Friedmann value even if the PBH density is small, thereby reducing the accretion.

Thus there are two distinct but related PBH constraints: one associated with the effects on the Universe's thermal history and the other with the generation of background radiation. This problem was first studied in Ref.~\cite{10.1093/mnras/194.3.639}. Even though this work neglected some factors and was later superseded by more detailed numerical investigations, we will use it here because it is the only analysis which applies for very large PBHs. The neglected factors mainly affect the time from which the analysis described below can be applied.

We assume that each PBH accretes at the Bondi rate
\begin{equation}
	\dot{M}
		\approx
					10^{11}
					\left(
						\frac{ M }{ M_{\odot} }
					\right)^{\!2}\.
					\left(
						\frac{ n }{ {\rm cm}^{-3} }
					\right)\!
					\left(
						\frac{ T }{ 10^{4}\,\Krm} 
					\right)^{\!-3/2}
					\grm\;\srm^{-1}
					\, ,
\end{equation}
where a dot indicates a derivative with respect to cosmic time $t$ and $n$ and $T$ are the particle number density and temperature of the gas at the black-hole accretion radius,
\begin{equation}
	R_{\arm}
		\approx
					10^{14}
					\left(
						\frac{ M }{ M_{\odot} }
					\right)\!
					\left(
						\frac{ T }{ 10^{4}\,\Krm }
					\right)^{\!-1}
					{\rm\,cm}
					\, .
\end{equation}
For large values of $M$, we will find that the Bondi formula may be inapplicable at early times and we return to this issue later. Each PBH will initially be surrounded by an HII region of radius $R_{\srm}$. If $R_{\arm} > R_{\srm}$ or if the whole Universe is ionised (so that the individual HII regions merge), the appropriate values of $n$ and $T$ are those in the background Universe ($\bar{n}$ and $\bar{T}$). In this case, after decoupling, $\dot{M}$ is epoch-independent so long as $\bar{T}$ has its usual Friedmann behaviour ($\bar{T} \propto z^{2}$). However, $\dot{M}$ decreases with time if $\bar{T}$ is boosted above the Friedmann value. If the individual HII regions have not merged and $R_{\arm} < R_{\srm}$, the appropriate values for $n$ and $T$ are those which pertain within the HII region. In this case, $T$ is usually close to $10^{4}\,$K and pressure balance at the edge of the region implies $n \sim \bar{n}\.( \bar{T} / 10^{4}\,\Krm )$. This implies $\dot{M} \propto z^{5}$, so the accretion rate rapidly decreases in this phase.

We assume that accreted mass is converted into outgoing radiation with constant efficiency $\epsilon$, so that the associated luminosity is
\vs{-1mm}
\begin{equation}
	L
		=
					\epsilon\,\dot{M}c^{2}
					\,
					,
\end{equation}
until this reaches the Eddington limit,
\begin{equation}
	L_{\rm ED}
		 = 
					\frac{ 4 \pi\.G M\.m_{\prm} }{ \sigma_{\Trm} }
		\approx
					10^{38}
					\left(
						\frac{ M }{ M_{\odot} }
					\right)
					{\rm erg}\,\srm^{-1}
					\, ,
\end{equation}
at which the Thomson drag of the outgoing radiation balances the gravitational attraction of the hole. (Here $\sigma_{\Trm}$ is the Thomson cross-section and $m_p$ is the proton mass.) The assumption that $\epsilon$ is constant may be simplistic and more sophisticated models allow it to be $\dot{M}$-dependent. We also assume that the spectrum of emergent radiation is constant, extending up to an energy $E_{\rm max} = 10\,\eta\,$keV, with Ref.~\cite{10.1093/mnras/194.3.639} considering models with $\eta = 0.01$, $1$ and $100$. This assumption is also simplistic but allows an analytic treatment of the problem.

We must distinguish between: (1) the local effect of a particular BH at distances sufficiently small that it dominates the effects of the others; and (2) the combined effect of all the BHs on the mean conditions of the background Universe. Both effects are very dependent on the spectrum of the accretion-generated radiation. As regards (1), the temperature in the HII region around each PBH is somewhat smaller than $10^{4}\,$K, falling as $\theta \approx ( z / 10^{3} )^{0.3}$, where $\theta$ is the temperature in units of $10^{4}\,$K, because the temperature inside the HII region is determined by the balance between photoionisation heating and inverse Compton cooling off the CMB photons. As regards (2), if the spectrum is hard, photons with energy much above $10\,$eV will escape from the HII region unimpeded, with most of the black-hole luminosity going into background radiation or global heating of the Universe through photoionisation when the background ionisation is low and Compton scattering off electrons when it is high. The matter temperature would generally be boosted well above its Friedmann value and, for a wide range of values for $M$ and $f_{\rm PBH}( M )$, the Universe would be reionised, which may itself be inconsistent with observations.

\begin{figure}[t]
	\includegraphics[width = 0.9\linewidth]{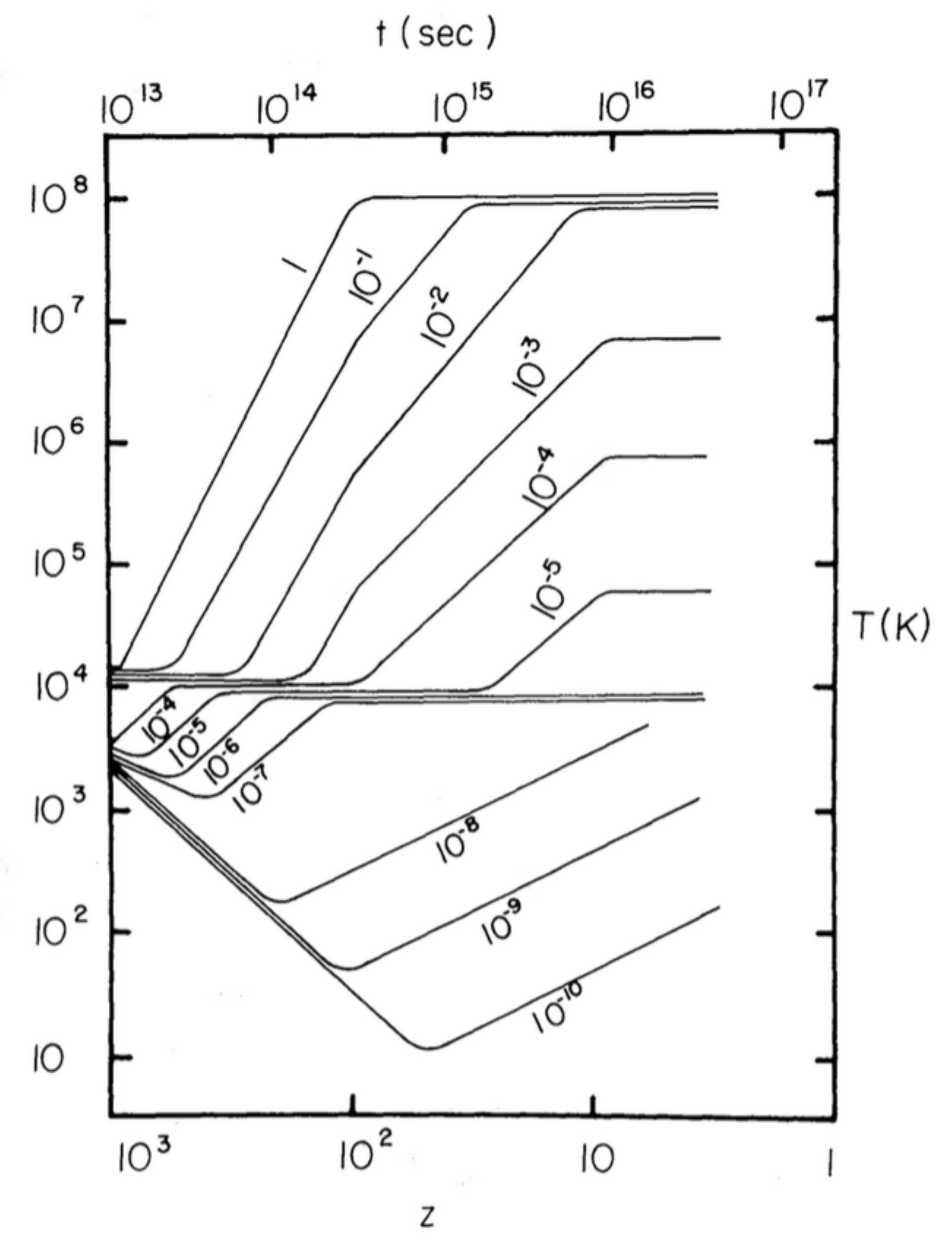} 
	\caption{This shows how the effect of PBH accretion on the evolution of 
			the matter temperature $T$ during the Eddington phase, 
			while independent of the mass $M$, depends on the PBH density $\Omega_{\rm PBH}$, 
			from Ref.~\cite{10.1093/mnras/194.3.639}.
			We assume that the energy of the accretion-generated photons is 
			$E_{\rm max} = 10\,$keV 
			and that $\Omega_{\grm} = 0.05$. For $\Omega_{\rm PBH} > 10^{-8}$, 
			$T$ will {\it always} deviate 
			from Friedmann behaviour and the whole Universe will be 
			collisionally re-ionised when 
			$T$ reaches $10^{4}\,$K. For $\Omega_{\rm PBH} > 10^{-4}$, 
			$T$ never falls below $10^{4}\,$K 
			(so the Universe does not go through a neutral phase 
			at all after decoupling) 
			and it will eventually rise above $10^{4}\,$K when 
			the Compton heating of the generated photons 
			exceeds the inverse Compton cooling of the CMB.
			$T$ flattens off when it reaches $T_{\rm max} \sim 10^{8}\,$K.}
	\label{fig:Tz1}
\end{figure}

Providing the Bondi formula applies, the analysis of Ref.~\cite{10.1093/mnras/194.3.639} shows that a PBH will accrete at the Eddington limit for some period after decoupling if
\begin{equation}
	M
		>
					M_{\rm ED}
		\approx
					10^{3}\.
					\epsilon^{-1}\.
					\Omega_{\grm}^{-1}\,
					M_{\odot}
					\, ,
\end{equation}
where $\Omega_{\grm} \approx 0.05$ is the gas density parameter, and this condition certainly applies for SLABs. This phase will persist until a redshift $z_{\rm ED}$ which depends upon $M$, $\epsilon$, $\Omega_{\grm}$ and $\Omega_{\rm PBH}$ and could be as late as galaxy formation for large enough PBHs. So long as the PBHs radiate at the Eddington limit, which could be true even if the Bondi formula is an overestimate, the evolution of the temperature in the background Universe is as indicated in Fig.~\ref{fig:Tz1}.

\begin{figure}[t]
	\includegraphics[width = 1.0\linewidth]{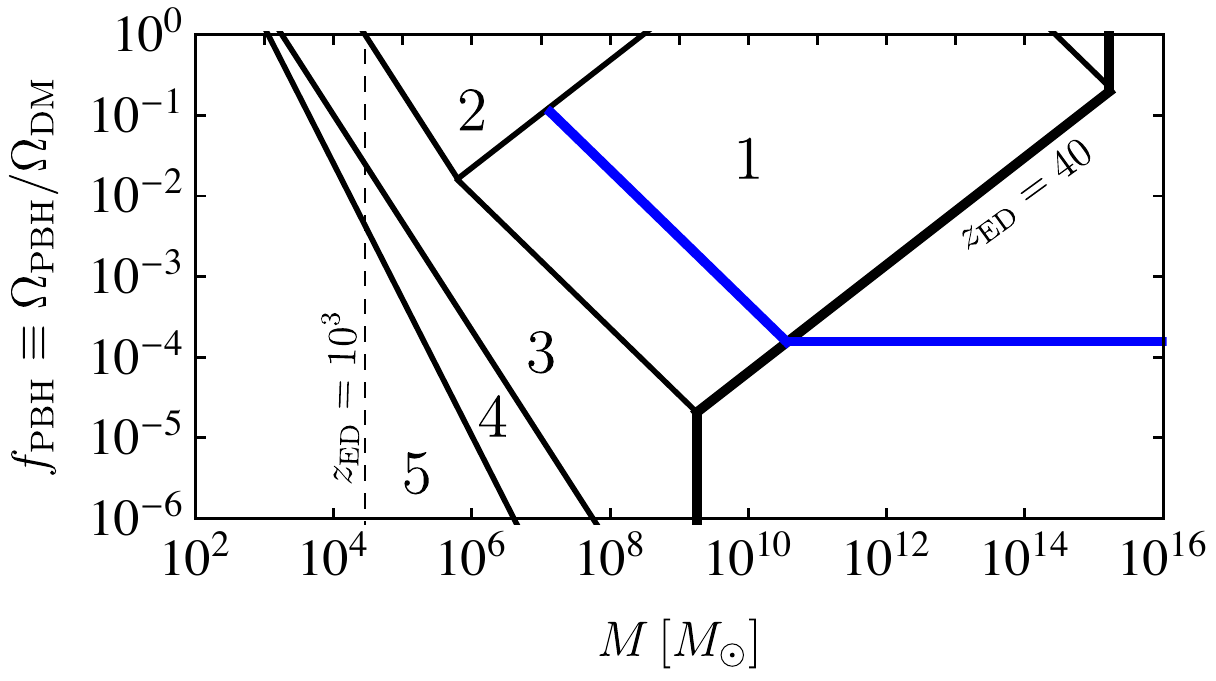} 
	\caption{
			This shows how the effect of PBH accretion on the evolution of 
			the background matter temperature depends on the PBH mass and density, 
			from Ref.~\cite{10.1093/mnras/194.3.639}.
			We assume $\epsilon = 0.1$, $\Omega_{\grm} = 0.05$ and $E_{\rm max} = 10\,$keV.
			The accretion rate exceeds the Eddington limit for some period after decoupling 
			to the right of the line $z_{\rm ED} = 10^{3}$ and the Eddington phase persists 
			throughout the pregalactic era to the right of the line $z_{\rm ED} = 10$.
			In each domain the end of the Eddington phase $t_{\rm ED}$ depends on 
			$\Omega_{\rm PBH}$ and $M$ in a different way.
			In domains (1) and (2), $T$ is boosted above $10^{4}\,$K by Compton heating; 
				$t_{\rm ED}$ exceeds $t_{*}$ in domain (1) but it is less than it in domain (2).
				Note that $T$ can attains the temperature of the hottest accretion-generated 
				photons above the line in the top right-hand corner of domain (1).
			In domain (3), $T$ is boosted up to $10^{4}\,$K but not above it and the whole 
				Universe is re-ionised, with no neutral phase at decoupling at all 
				for $\Omega_{\rm PBH} > 10^{-4}$. The ionised phase necessarily persists 
				until galaxy formation in domains (1) and (2); 
				it may also do so in parts of domain (3).
			In domain (4), the Universe is not re-ionized but there is a period 
				in which $T$ rises.
			In domain (5), $T$ always falls but, for a period after decoupling, 
				it stays at the CMB temperature rather than falling like $z^{2}$.}
	\label{fig:history}
\end{figure}

The overall effect on the thermal history of the Universe (\ie~going beyond the Eddington phase) for different ($\Omega_{\rm PBH},\,M$) domains is indicated in Fig.~\ref{fig:history}. Again this assumes the Bondi formula applies.
	In domain (1), $\bar{T}$ is boosted above $10^{4}\,\Krm$ and the Eddington phase persists until after the redshift~\cite{10.1093/mnras/194.3.639}
\begin{equation}
	\label{eq:zstar-def}
	z_{*}
		\approx
					10^{2}\,\Omega_{\mrm}^{1/3}\.
					(
						\eta\,\Omega_{\grm}
					)^{-2/3}
					\, 
\end{equation}
at which most of the black-hole radiation goes into Compton heating; $\bar{T}$ is boosted to $E_{\rm max}$ in the top right.
	In domain (2), $\bar{T}$ is also boosted above $10^{4}\,$K but the Eddington phase ends before $z_{*}$. 
	In domain (3), $\bar{T}$ is boosted to $10^{4}\,$K but not above it because of the cooling of the CMB. 
	In domain (4), $\bar{T}$ increases for a while but does not reach $10^{4}$K, so the Universe is not re-ionised. 
	In domain (5), $\bar{T}$ never increases during the Eddington phase but follows the CMB temperature for a while (\ie~falls like $z$). We are mainly interested in domains (1) to (3) and in these the Eddington phase ends at a redshift~\cite{10.1093/mnras/194.3.639}
\begin{equation}
	\label{eq:zED}
	z_{\rm ED}
		\approx
					\begin{cases}
						10^{3.8}\.
						(
							\Omega_{\rm PBH}\.\eta
						)^{2/9}\.
						(
							M \epsilon\.\Omega_{\grm} / M_{\odot}
						)^{-4/27}
							& ( 1 )
						\, ,
						\\[2.5mm]
						10^{3.3}\.\Omega_{\rm PBH}^{1/6}\.
						(
							M \epsilon / M_{\odot}
						)^{-1/9}\.
						\Omega_{\grm}^{-5/18}
							& ( 2 )
						\, ,
						\\[2.5mm]
						10^{4.0}\.
						(
							M \epsilon\,\Omega_{\grm} / M_{\odot}
						)^{-1/3} 
							& ( 3 )
						\, .
					\end{cases}
\end{equation}

Following Ref.~\cite{10.1093/mnras/189.1.123}, we now derive constraints on the PBH density by comparing the time-integrated emission to the observed background intensity. In domain (1) the maximum contribution comes from the end of the Eddington phase ($z_{\rm max} = z_{\rm ED}$), which is after the epoch $z_{*}$; in domain (2) it comes from the epoch $z_{\rm max} = z_{*}$, somewhat after the end of the Eddington phase; in domain (3) it also comes from this epoch but the background temperature never goes above $10^{4}\,$K and the background light limit turns out to be unimportant. The redshifted time-integrated energy production per PBH in the relevant domains is 
\begin{equation}
	E
		\approx
					\begin{cases} 
						10^{46}\.
						( \epsilon\,\Omega_{\grm} )^{10/27}
						( \Omega_{\rm PBH}\.\eta )^{-5/9}
						\left(
							M / M_{\odot}
						\right)^{37/27}
						\!{\rm erg} 
							& \!\!\!\!\!(1)
						\, ,
							\\[1.5mm]
						10^{45.5}\.\epsilon^{2/5}\.
						\Omega_{\rm PBH}^{3/5}\.
						\eta^{-11/15} \,\Omega_{\grm}^{4/15}
						\left(
							M / M_{\odot}
						\right)^{7/5}
						\!{\rm erg}
							& \!\!\!\!\!(2)
						\, .
					\end{cases}_{_{_{_{_{}}}}}
\end{equation}
If $\eta = 1$, corresponding to $E_{\rm max} = 10\,$keV, then in domain (1), where $z_{\rm max} \sim ( 10$ -- $100 )$, the radiation would reside in the range $(0.1$ -- $1)\,$keV where the observed background radiation density is $\Omega_{\Rrm} \sim 10^{-7}$; in domain (2), where $z_{\rm max} \sim 100$, the radiation would presently reside at $\sim 100\,$eV where $\Omega_{\Rrm} \sim 10^{-6.5}$. The associated limit on the PBH density parameter is then
\begin{widetext}
\begin{equation}
	\label{eq:accretion}
	\Omega_{\rm PBH}
		<
					\begin{cases} 
						( 10\.\epsilon )^{-5/6}\.
						\big(
							M / 2 \times 10^{5}\,M_{\odot}
						\big)^{-5/6}\.
						\eta^{5/4}\.
						\big(
							\Omega_{\grm} / 0.05
						\big)^{-5/6}
							& (1)
						\, ,
						\\[1.5mm]
						( 10\.\epsilon )^{-1}\.
						\big(
							M / 3 \times 10^{4}\,M_{\odot}
						\big)^{-1}\.
						
							\eta^{11/6} \. \big( \Omega_{\grm} / 0.05
						\big)^{7/6}
							& (2)
						\, .
					\end{cases}_{_{_{_{_{}}}}}
\end{equation}
\end{widetext}
There is a discontinuity at the 1/2 boundary because of the assumed jump in $\Omega_{\Rrm}$. Clearly this is unrealistic since $\Omega_{\Rrm}$ would vary continuously across the ($f_{\rm PBH}, M$) plane in a more precise analysis. We therefore only show the domain (1) limit in Fig.~\ref{fig:history}, as indicated by the upper part of the blue line. This constraint depends on the validity of the Bondi formula with constant $M$, so we now consider whether this is applicable.

Three feature could modify the above analysis. First, the accretion which generates the luminosity also increases the BH mass, so we need to consider the consequences of this. During the Eddington phase, each BH doubles its mass on the Salpeter timescale, $t_{\Srm} \approx 4 \times 10^{8}\.\epsilon$\,yr~\cite{Salpeter:1964kb}, so $M$ can only be regarded as constant if $t_{\rm ED} < t_{\Srm}$; this implies $z_{\rm ED} > z_{\Srm} \approx 10\,\epsilon^{-2/3}$. From Eq.~\eqref{eq:zED}, this corresponds to ($\Omega_{\rm PBH}, M$) values to the left of the bold line in Fig.~\ref{fig:history}, which is given by 
\vs{-1mm}
\begin{equation}
	\label{eq:Mconst}
	M
		<
					\begin{cases}
						10^{19}\,
						\Omega_{\grm}^{-1}\.
						\epsilon^{7/2}\.
						( \Omega_{\rm PBH}\.\eta )^{3/2}\,
						M_{\odot}
							& ( 1 )
						\, ,
						\\[1.5mm]
						10^{9}\,
						\Omega_{\grm}^{-1}\.
						\epsilon\,
						M_{\odot}
							& ( 3 )
						\, .
					\end{cases}
\end{equation}
To the right of the bold line, as may apply for SLABs, the PBH mass increases by a factor
\vs{-1mm}
\begin{equation}
	\label{eq:salpetergrowth}
	M / M_{i}
		\approx
					\exp( t_{\rm ED} / t_{\Srm} )
		\approx
					\exp
					\big[
						(0.1/ \epsilon)( 40 / z_{\rm ED} )^{3/2}
					\big]
					\, ,
\end{equation}
where $M_{i}$ is the {\it initial} mass of the PBH, so the previous analysis is inconsistent in this region. It is therefore best to regard $M$ in Eq.~\eqref{eq:zED} and Fig.~\ref{fig:history} as the {\it current} mass and accept that the above analysis only applies above the bold line. However, one can analyse the problem more carefully by using Eq.~\eqref{eq:salpetergrowth} to express Eq.~\eqref{eq:zED} in terms of $M_{i}$. One then finds a limiting initial mass above which the Eddington phase extends right up to galaxy formation, when the model breaks down anyway.

The constraint in Eq.~\eqref{eq:accretion} does not apply below the bold line in Fig.~\ref{fig:history} and it intersects this at 
\vs{-1mm}
\begin{equation}
	M
		\approx
					10^{9}\.( 10\.\epsilon )\.
					\eta^{3/2}\.
					\Omega_{\grm}^{-1}\,M_{\odot}
					\, .
\end{equation}
However, the following 
argument gives its form for larger $M$. Since most of the final black-hole mass generates radiation with efficiency $\epsilon$, the current energy of the radiation produced is $E( M ) \approx \epsilon\.M c^{2} / z( M )$ where the redshift at which the radiation is emitted must satisfy $z( M ) < 10\.\epsilon^{-2/3}$. The current background radiation density is therefore $\Omega_{\Rrm} \approx \epsilon\,\Omega_{\rm PBH} z( M )^{-1}$, so the constraint becomes
\vs{-2mm}
\begin{equation}
	\label{eq:accretion2}
	\Omega_{\rm PBH}
		<
					\epsilon^{-1}\.\Omega_{\Rrm} \. z_{\Srm}
		\approx
					10^{-5}\.( 10\.\epsilon )^{-5/3}
					\, .
\end{equation} 
This limit is shown by the flat part of the blue line in Fig.~\ref{fig:history}. This is equivalent to the well-known Soltan constraint~\cite{Soltan:1982vf} from observations of the X-ray background on the SMBHs that power quasars. A more precise calculation would be required to derive the exact transition from the limit~\eqref{eq:accretion}. The evolution of the PBH mass and temperature after $t_{\rm ED}$ is complicated in this case but 
limit~\eqref{eq:accretion2} is independent of this.

\vs{2mm}

\begin{figure}[t]
	\includegraphics[width = 0.9\linewidth]{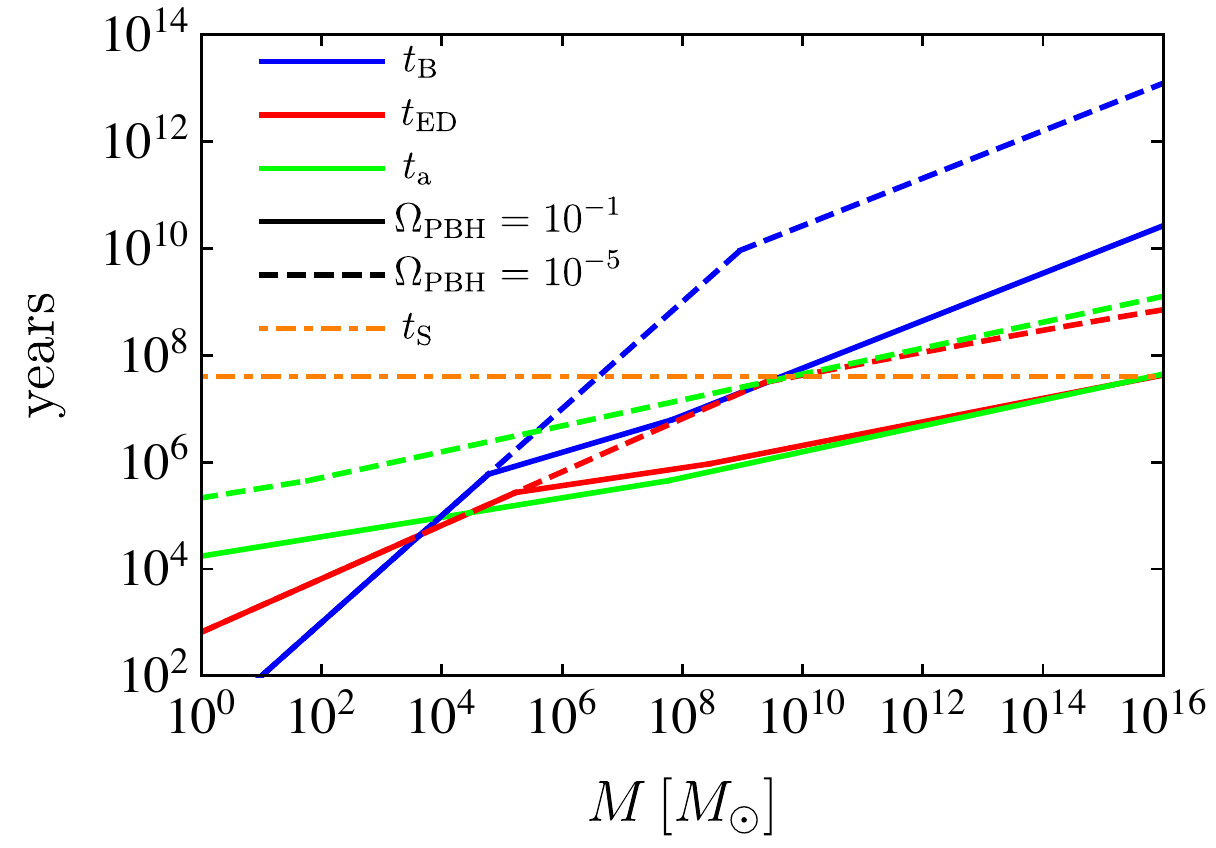}
	\caption{The end of the Eddington phase $t_{\rm ED}$ (red), the Bondi timescale at that epoch 
		$t_{\Brm}( t_{\rm ED} )$ (blue) and the time at which radiation drag becomes unimportant
		$t_{\arm}$ (green) as functions of $M$ for $\Omega_{\rm PBH} = 0.1$ (solid) and $10^{-5}$ (broken).
		The BH increases it mass appreciably if $t_{\rm ED}$ exceeds the Salpeter timescale $t_{\Srm}$ 
		(black dotted). The steady-state Bondi formula is applicable for $t_{\rm ED} < t_{\Brm}( t_{\rm ED} )$ 
		and radiation drag can be neglected by the end of the Eddington phase for $t_{\rm ED} > t_{\arm}$, 
		both conditions applying for sufficiently small $M$.}
	\label{fig:zED}
\end{figure}
\vs{2mm}

The second problem is that the steady-state assumption fails if the Bondi accretion timescale, 
\begin{equation}
	\label{eq:steady}
	t_{\Brm}
		\approx
					10^{12}
					\(
						\frac{ M }{ 10^{4}\,M_{\odot} }
					\)\!
					\(
						\frac{ T }{ 10^{4}\,\Krm }
					\)^{\!-3/2}
					\,\srm
					\, ,
\end{equation}
exceeds the cosmic expansion time (\ie~the Bondi formula is inapplicable at times earlier than $t_{\Brm}$). This is equivalent to the condition that the mass within the accretion radius exceeds $M$, a problem already discussed in Ref.~\cite{10.1093/mnras/194.3.639}. For $R_{\arm} < R_{\srm}$ or in domain (3), one has $T \approx 10^{4}$\,K, so Eq.~\eqref{eq:steady} implies that the Bondi formula applies only at times later than $10^{8}\.( M / M_{\odot} )\,$s. In domains (1) and (2), $T$ is increased, so the calculation of $t_{\Brm}$ is more complicated. Ref.~\cite{10.1093/mnras/194.3.639} argues that the large mass within the accretion radius will complicate the dynamics of the accretion flow but that there is no reason for supposing that $\dot{M}$ will be reduced relative to the Bondi rate. However, that conclusion is questionable and probably the accretion radius is reduced to the value within which the gas mass contained is comparable to $M$~\footnote{The neglect of the cosmic expansion also invalidates the use of the Bondi formula during the radiation era and this hugely reduces the expected PBH accretion~\cite{Carr:1974nx}. The consequences in the matter era may be less significant but are still uncertain.}. In fact, the situation is analogous to that discussed in Ref.~\cite{Yazdi:2016fyt} in the context of AGN accretion, where fragmentation into stars is assumed to reduce the accretion.

Because of the uncertainty, we now consider the steady-state condition more carefully. Clearly Eq.~\eqref{eq:zED} for $z_{\rm ED}$ applies only if this corresponds to a time later than $t_{\Brm}$ at that epoch. Since Eq.~\eqref{eq:zED} implies that $t_{\rm ED} \propto z_{\rm ED}^{-3/2}$ increases more slowly than $M$, the steady-state condition can only be satisfied if $T$ increases above $10^{4}\,$K, which requires that one be in domain (1) or (2). The temperature in these domains is determined by the balance of Compton heating from accretion-generated photons and Compton cooling off the CMB and this gives~\cite{10.1093/mnras/194.3.639}
\begin{equation}
	T
		\approx
					\begin{cases}
						10^{16}\,\Omega_{\rm PBH}\,
						\Omega_{\grm}^{-1}\,z^{-4}\,\Krm
							& ( z > z_{*} )
						\, ,
						\\[1.5mm]
						10^{13}\,\Omega_{\rm PBH}\,\eta\,
						\Omega_{\mrm}^{-1/2}\,z^{-5/2}\.\Krm
							& ( z < z_{*} )
						\, ,
							\\
					\end{cases}
\end{equation} 
where $z_{*}$ is defined by Eq.~\eqref{eq:zstar-def}. Setting $z = z_{\rm ED}$ in this expression then gives $t_{\Brm}$ at $t_{\rm ED}$ as a function of $M$ and $\Omega_{\rm PBH}$. This is plotted in Fig.~\ref{fig:zED} and compared with the function $t_{\rm ED}( M )$ for particular values of $\Omega_{\rm PBH}$. This shows that the Bondi formula is applicable at $t_{\rm ED}$ only for $M < 10^{4}\,M_{\odot}$.

We conclude that the increase in the background temperature does not suffice to restore steady-state accretion before the end of the Eddington phase for SLABs. However, the implications of this remain unclear. Possibly one might expect the solution to be described by self-similar infall instead~\cite{Bertschinger:1985pd}. Figure~\ref{fig:zED} also shows the Salpeter timescale $t_{\Srm}$, so the BH mass increases appreciably only where this falls below the $t_{\rm ED}$ line (\ie~only for very large values of $M$).

The third problem is that accreting gas will have an inward velocity ($v_{\rm in}$) relative to the expanding background of CMB photons and the Thomson drag of these photons will inhibit accretion at sufficiently early times. If the drag per particle ($\sim \rho_{\Rrm}\.\sigma_{\Trm}\.c\.v_{\rm in}$ where $\rho_{\Rrm}$ is the radiation density) exceeds the gravitational attraction of the hole at the accretion radius ($G M m_{\prm} / R_{\arm}^{2}$), the {\it effective} accretion radius will be reduced to
\begin{equation}
	R_{*}
		\approx
					\left(
						\frac{ G M\.m_{\prm}\.t }
						{ \rho_{\Rrm}\.\sigma_{\Trm}\.c }
					\right)^{\!1/3}
					\, ,
\end{equation}
where the drag and attraction balance. This implies that accretion is reduced until the time at which $R_*$ reaches $R_{\arm}$. If $R_{\arm} < R_{\srm}$ or in domain (3), one can assume $T \approx 10^{4}\,$K and this time can be shown to be
\begin{equation}
	t_{\arm}
		\approx
					\begin{cases} 
						10^{11}\,( M / M_{\odot} )^{3/8}\,
						\Omega_{\mrm}^{-1/2}\,\srm
							& \!\!\!\!( 10^{6} < M / M_{\odot} < 10^{7} )
						\, ,
							\\[1.5mm]
						10^{10}\,( M / M_{\odot} )^{6/11}\,
						\Omega_{\mrm}^{-4/11}\,\srm
							& \!\!\!\!( M > 10^{7} \, M_{\odot} )
						\, ,
					\end{cases}
\end{equation} 
The mass scales $M \sim 10^{6}\,M_{\odot}$ and $10^{7}\,M_{\odot}$ correspond to $t_{\arm} \sim 10^{13}\,$s (decoupling) and $t_{\arm} \sim 10^{14}\,$s, respectively. In domains (1) and (2), one must account for the temperature increase and one finds
\begin{equation}
	t_{\arm}
		\approx
					\begin{cases}
						4 \times 10^{3}
						\(
							M / M_{\odot}
						\)^{3/13}
						\(
							\eta\,\Omega_{\rm PBH}
						\)^{-9/26}
						{\rm \,yr}
							& (1)
						\, ,
						\\[1.5mm]
						3\times 10^{4}
						\(
							M / M_{\odot}
						\)^{6/35}\.
						(
							\Omega_{\grm} / \Omega_{\rm PBH}
						)^{9/35}\,
						{\rm \,yr}
							& (2)
						\, .
					\end{cases}
\end{equation} 
This function is also shown in Fig.~\ref{fig:zED} and the expression for the end of the Eddington phase is unaffected if $t_{\rm ED} > t_{\arm}$ . We can see that radiation drag is already unimportant at decoupling for $M < 10^{6}\,M_{\odot}$ and that it becomes unimportant before the end of the Eddington phase for all $M$. So the only effect of the drag is to postpone the onset of this phase.

\subsection{More recent accretion studies}

Later an improved numerical analysis of pregalactic PBH accretion was provided by Ricotti {\it et al.}~\cite{Ricotti:2007au}. They used a more realistic model for the efficiency parameter $\epsilon$, allowed for the increased density in the dark halo expected to form around each PBH and included the effect of the velocity dispersion of the PBHs on the accretion in the period after cosmic structures start to form. They found much stronger accretion limits by considering the effects of the emitted radiation on the spectrum and anisotropies of the CMB rather than the background radiation itself. Using FIRAS data to constrain the first, they obtained a limit $f_{\rm PBH}( M ) < ( M / \Msun )^{-2}$ for $1\,\Msun < M \lesssim 10^{3}\,\Msun$; using WMAP data to constrain the second, they obtained a limit $f_{\rm PBH}( M ) < ( M / 30\,\Msun )^{-2}$ for $30\,\Msun < M \lesssim 10^{4}\,\Msun$. The constraints flatten off above the indicated masses but are taken to extend up to $10^{8}\,\Msun$. Although these limits appeared to exclude $f_{\rm PBH} = 1$ down to masses as low as $1\,\Msun$, they were very model-dependent and there was also a technical error (an incorrect power of redshift) in the calculation.

This problem has been reconsidered by several groups, who argue that the limits are weaker than indicated in Ref.~\cite{Ricotti:2007au}. Ali-Ha{\"i}moud and Kamionkowski~\cite{Ali-Haimoud:2016mbv} calculate the accretion on the assumption that it is suppressed by Compton drag and Compton cooling from CMB photons and allowing for the PBH velocity relative to the background gas. They find the spectral distortions are too small to be detected, while the anisotropy constraints only exclude $f_{\rm PBH} = 1$ above $10^{2}\,\Msun$. Horowitz~\cite{Horowitz:2016lib} performs a similar analysis and gets an upper limit of $30\,\Msun$. Poulin {\it et al.}~\cite{Poulin:2017bwe, Serpico:2020ehh} argue that the spherical accretion approximation probably breaks down, with an accretion disk forming instead, and this affects the statistical properties of the CMB anisotropies. Provided the disks form early, these constraints exclude a monochromatic distribution of PBH with masses above $2\,\Msun$ as the dominant form of dark matter. However, none of these analyses considers masses above $10^{4}\,M_{\odot}$, which is why we have focussed on the old analysis of Ref.~\cite{10.1093/mnras/194.3.639}.

\section{SLAB Constraints from WIMP Annihilations}
\label{sec:SLAB-Constraints-from-WIMP-Annihilations}

If PBHs do not constitute {\it most} of the dark matter, the question of the nature of the remaining part arises. In the following analysis, we assume a mixed DM scenario in which the PBHs are subdominant, \ie~$f_{\rm PBH} \equiv \rho_{\rm PBH} / \rho_{\rm DM} \ll 1$, where $\rho_{\rm DM}$ is the total observed DM energy density. In more detail, the dominant dark-matter component is taken to be a WIMP, whose abundance is set through a thermal mechanism, although our conclusions below hold in more general cases. We therefore assume a WIMP density $\rho_{\chi} \equiv f_{\chi}\.\rho_{\rm DM}$ with $f_{\chi} + f_{\rm PBH} = 1$. One could envisage a scenario in which $f_{\rm PBH}$ is very small for SLABs but close to unity in some other mass range, in which case both $f_{\rm SLAB}$ and $f_{\chi}$ could be small. We discuss such a scenario in an accompanying paper~\cite{Visinelli:2020B}.

\subsection{Structure of the dark-matter halos}

We first consider the production of the WIMP number density through thermal freeze-out. When the annihilation rate falls below the expansion rate of the Universe, the number of WIMPs per comoving volume freezes out. This occurs at a temperature given by $k_{\Brm}\.T^{}_{\Frm}\! \sim m_{\chi}\.c^{2} / 20$~\cite{Lee:1977ua, Steigman:1979kw}, where $m_{\chi}$ is the mass of the WIMP and $k_{\Brm}$ is Boltzmann's constant. Even after freeze-out, the relativistic plasma and WIMP population keep exchanging energy and momentum until the {\it scattering} rate falls below the Hubble rate at kinetic decoupling (KD)~\cite{Bernstein:1985th}. This leaves an imprint on the current WIMP velocity dispersion, since the photon temperature $T_{\gamma}$ scales as $a^{-1}$ after KD, while the temperature of the non-relativistic WIMPs scales as $a^{-2}$. We use the following expression for the KD temperature~\cite{Bringmann:2006mu}:
\begin{equation}
	k_{\Brm}\.T_{\rm KD}
		=
					\frac{ m_{\chi}\.c^{2} }{ \Gamma( 3 / 4 ) }\mspace{-2mu}
					\(
						\frac{ g\,m_{\chi} }{ \MP }
					\)^{\!1/4}
					\, ,
\end{equation}
where $g \approx 10.9$ for temperatures in the range $0.1$ -- $10\,$MeV and $\Gamma( 3 / 4 ) \approx 1.225$. This expression coincides, within a numerical factor, with other definitions in the literature~\cite{Visinelli:2015eka}. The corresponding Hubble rate and time are $H_{\rm KD}$ and $t_{\rm KD} = 1 / ( 2\.H_{\rm KD} )$, respectively.

PBHs are formed prior to $t_{\rm eq}$ (\ie~during the radiation-dominated epoch) from the direct collapse of mildly non-linear perturbations. The PBH mass at KD is
\begin{equation}
	\label{eq:turnaround-rg}
	M_{\rm KD}
		\approx
					300\,M_{\odot}\,m_{100}^{\!5/4}
					\, ,
\end{equation}
where $m_{100} = m_{\chi}\,c^{2}/(100\,{\rm GeV})$. After PBH formation, the WIMPs will be gravitationally attracted to the PBHs, leading to the formation of surrounding halos. The structure of these halos depends on the specific circumstances and particle velocities. The fraction of WIMPs with low velocities remain gravitationally bound to the PBHs and form density spikes around them. For PBHs smaller than $M_{\rm KD}$, the WIMP density is expected to be uniform, since these PBHs have formed before kinetic decoupling, when the WIMPs are still tightly coupled to the plasma. WIMP accretion occurs during two different periods:
	(1) between the kinetic decoupling and $t_{\rm eq}$;
	(2) through secondary accretion after $t_{\rm eq}$.
The halo mass is never much more than the PBH mass in the first phase but it can be much larger in the second phase. In both cases, the WIMPs form a halo with a universal density profile $\rho_{\chi}( r ) \propto r^{-9/4}$ and this halo grows with time~\cite{Bertschinger:1985pd}. We elaborate on this and the mass ranges involved below.

\subsection{Formation of WIMP halo around PBH}

WIMPs which are non-relativistic after freeze-out can form a gravitationally-bound halo around PBHs~\cite{Lacki:2010zf, Saito:2010ts, Xu:2020jpv} immediately after kinetic decoupling~\cite{Eroshenko:2016yve, Boucenna:2017ghj, Eroshenko:2019pxt, Adamek:2019gns}. Assuming comoving entropy conservation, the WIMP energy density at a time $t < t_{\rm eq}$ is
\begin{equation}
	\label{eq:rhoDM}
	\rho_{\rm \chi\,spike}
		 = 
					f_{\chi}\.\frac{ \rho_{\rm eq} }{ 2 }\mspace{-2mu}
					\left(
						\frac{ a }{ a_{\rm eq} }
					\right)^{\!-3}
		\approx
					f_{\chi}\.\frac{ \rho_{\rm eq} }{ 2 }\mspace{-2mu}
					\left(
						\frac{ t }{ t_{\rm eq} }
					\right)^{\!-3/2}
					,
\end{equation}
where $\rho_{\rm eq} = 3 / ( 32 \pi\.G\.t_{\rm eq}^{2} )$. The second expression neglects any change in the entropy degrees of freedom. In order to find the extent of the DM profile around a PBH, we consider the turn-around point of the radial motion of an orbiting particle, assuming the Newtonian equation,
\begin{equation}
	\label{eq:motion}
	\ddot r
		 = 
					-
					\frac{ GM }{ r^{2} }
					+
					\frac{ \ddot a }{ a }\.r
		 = 
					-
					\frac{ GM }{ r^{2} }
					-
					\frac{ r }{ 4\.t^{2} }
					\, , 
\end{equation}
where $r$ is the distance of the particle from the PBH. The second expression holds for a radiation-dominated universe with $a \propto t^{1/2}$. The numerical solution for the turn-around radius obtained from Eq.~\eqref{eq:motion} is well approximated by~\cite{Adamek:2019gns}
\begin{equation}
	\label{eq:turnaround}
	r_{\rm ta}( t )
		\approx
					\left[
						r_{\grm}\.( c\.t )^{2}
					\right]^{1/3}
					\, ,
\end{equation}
where $r_{\grm} = 2\.G M / c^{2}$. This is just the condition that the two terms on the right-hand-side of Eq.~\eqref{eq:motion} are comparable. As explained in Ref.~\cite{Adamek:2019gns}, Eq.~\eqref{eq:turnaround} is just the evolving radius within which the cosmological mass is comparable to the PBH mass since overdensities do not grow during the radiation era.

Using Eqs.~\eqref{eq:rhoDM} and~\eqref{eq:turnaround}, it can be shown that the density profile of the WIMPs around the BH at time $t_{\rm eq}$ corresponds to a spike with~\cite{Adamek:2019gns}
\begin{eqnarray}
	\label{eq:densityprofile}
	\rho_{\rm \chi\,spike}( r ) 
		 &=& 
					f_{\chi}\.\frac{ \rho_{\rm eq} }{ 2 }\!
					\left(
						\frac{ r_{\rm ta}( t_{\rm eq} ) }{ r }
					\right)^{\!9/4}
					\notag
					\\[1mm]
		 &=&
					f_{\chi}\.\frac{ \rho_{\rm eq} }{ 2 }\!
					\left(
						\frac{ M }{\,M_{\odot} }
					\right)^{\!3/4}
					\left(
						\frac{ r_{\orm} }{ r }
					\right)^{\!9/4}
					\, ,
\end{eqnarray}
where $r_{\orm} = ( 2\.G\,M_{\odot}\.t_{\rm eq}^{2} )^{1/3} = 0.0193\,{\rm pc}$. This just comes from the cosmological density at the epoch when $r$ is the turn-around radius. Eq.~\eqref{eq:densityprofile} only applies up to the radius $r_{\rm ta}( t_{\rm eq} )$ and the mass within this radius is comparable to $M$ (as expected). This profile has been confirmed by numerical simulations for PBHs of $30\,M_{\odot}$~\cite{Adamek:2019gns} but it should also hold for more massive PBHs. In principle, the orbital motion of the WIMPs would influence their density profile~\cite{Eroshenko:2016yve}. However, the WIMP kinetic energy can be neglected for the PBH masses relevant to this paper. In this regime, a detailed derivation of the WIMP density profile after KD leads to the density profile~\eqref{eq:densityprofile} multiplied by a concentration parameter $\alpha_{\Erm} \approx 1.53$~\cite{Adamek:2019gns}.

After matter-radiation equality, the mass gravitationally bound by the PBH grows according to
\begin{equation}
	\tilde{M}( z )
		=
					M
					\left(
						\frac{ 1 + z_{\rm eq} }
						{ 1 + z }
					\right)
					\,
\end{equation}
and this is described as `secondary' accretion~\cite{Sikivie:1996nn}. Equivalently, the overdensity on a scale $\tilde{M}$ is $M / \tilde{M}$ at $t_{\rm eq}$, so the mass binding at $t$ is
\begin{equation}
	\tilde{M}
		\sim
					M\.
					(
						t / t_{\rm eq}
					)^{2/3}
		\sim
					M\.
					(
						\rho_{\rm eq} / \rho
					)^{1/3}
					,
\end{equation}
corresponding to radius 
\begin{equation}
	r
		\sim
					(
						\tilde{M} / \rho
					)^{1/3}
		\sim
					M^{1/3}\.
					\rho_{\rm eq}^{1/9}\.
					\rho^{-4/9}
					\, ,
\end{equation}
which just gives $\rho \propto r^{-9/4}$. Therefore, at a given redshift, secondary infall and virialisation lead to a DM density spike with the same radial dependence as Eq.~\eqref{eq:densityprofile}. This is confirmed by the numerical calculations of Ref.~\cite{Adamek:2019gns}. The accretion halts around the epoch of galaxy formation, which we set at $z_{\star} \sim 10$, because of the effects of dynamical friction between DM halos and hierarchical structure formation.

The WIMP population inside the halo is consumed by self-annihilation~\cite{Berezinsky:1992mx}. This gives a maximum concentration at redshift $z$ of
\begin{equation}
	\label{eq:rhomax}
	\rho_{\rm \chi\,max}( z )
		 = 
					f_{\chi}\.
					\frac{ m_{\chi}\,H( z ) }{ \sv }
					\, ,
\end{equation}
where $\sv$ is the velocity times the WIMP self-annihilation cross-section, averaged over the velocity distribution, and $H( z ) = H_{0}\,h( z )$ is the Hubble rate at redshift $z$. Combining these results gives the WIMP profile
\begin{equation}
	\label{eq:dmdistribution}
	\rho_{\chi}( r )
		 = 
					\min\!
					\big[ 
						\rho_{\rm \chi\,max}( z ),\,
						\alpha_{\Erm}\,\rho_{\rm \chi\, spike}( r )
					\big]
					\, .
\end{equation}
The extent of the plateau $r_{\rm cut}$ is obtained by equating the two expressions in Eq.~\eqref{eq:dmdistribution},
\begin{equation}
	\label{eq:rcut}
	r_{\rm cut}( z )
		\approx
					21\.
					\big[
						m_{100}\.h( z )
					\big]^{- 4 / 9}\!
					\left(
						\frac{ M }{ 10^{10}\,M_{\odot} }
					\right)^{\!\!1 / 3}	
					{\rm\,pc}		
					\, .
\end{equation}

\subsection{WIMP annihilation rate around PBHs}

We assume that WIMPs annihilate into Standard Model particles, in particular photons, through the s-wave channel, with no significant contribution from co-annihilation~\cite{Griest:1990kh} or Sommerfeld enhancement~\cite{ArkaniHamed:2008qn}. The thermal freeze-out mechanism fixes the number of WIMPs in a comoving volume for a specific value of the velocity-average WIMP annihilation cross-section $\sv$~\cite{Lee:1977ua, Hut:1977zn, Sato:1977ye}. The value of $\sv$ is independent on the WIMP velocity distribution to lowest order in $v/c$ and, for non-relativistic WIMPs, it is the same throughout the history of the Universe. Here, we set $\sv = \sv^{}_{\Frm} / f_{\chi}$, where $\sv^{}_{\Frm} \approx 3 \times 10^{-26}\,{\rm cm^{3}\,s^{-1}}$ is the value of $\sv$ required for WIMPs with $m_{\chi} \gtrsim 10\,$GeV to be produced at thermal freeze-out with $f_{\chi} = 1$. The scaling $\sv \propto 1/ f_{\chi}$ is expected within the standard freeze-out theory~\cite{Baum:2016oow}.

The annihilation rate is taken to be
\begin{align}
	\Gamma( z )
		&\equiv
					\frac{ \sv }{ m_{\chi}^{2} }
					\int\!\d r\,4\pi r^{2}\,
						\rho_{\chi}^{2}( r )
						\notag
						\\[1.5mm]
		 &= 
					\frac{ 4\pi\.\sv }{ m_{\chi}^{2} }\,
						\rho_{\rm \chi\,max}^{2}( z )\,
					r_{\rm cut}^{3}( z )
					\, ,
\end{align}
where the last expression assumes the DM density profile given by Eq.~\eqref{eq:dmdistribution}. The redshift dependence of the decay rate can be expressed as 
\vs{-1mm}
\begin{align}
	\Gamma( z )
		=
					f_{\chi}^{5/3}\,\Gamma_{0}\,
					[ h( z ) ]^{2/3}
					\, ,
\end{align}
where
\vs{-1mm}
\begin{equation}
	\label{eq:Gamma}
	\Gamma_{0}
		\approx
					\frac{ 3 }{ 8 }\mspace{-1mu}
					\left(
						\frac{ \alpha_{\Erm}^{4}\,\sv^{}_{\Frm}\,
						\rho_{\rm eq}\,H_{0}^{2} }
						{ 2\.m_{\chi}^{4} }
					\right)^{\!1/3}
					M
		\equiv
					\Upsilon\.M
					\,
\end{equation}
and $\Upsilon$ has units of ${\rm g^{-1}\,s^{-1}}$.

\subsection{Extragalactic background flux}

The extragalactic gamma-ray flux produced by the collective annihilations around PBHs at all redshifts $z$ is~\cite{Ullio:2002pj}
\begin{equation}
	\label{eq:flux}
	\frac{ \d\Phi_{\gamma} }{ \d E\.\d\Omega }\bigg|_{\rm e.g.}
		 \!\!\!\!=\! 
					\frac{ c }{ 8\pi}
					\int_{0}^{\infty}\!\d z\,
						\frac{ e^{-\tau_{\Erm}(z,\.E)} }{ H( z ) }
						\frac{ \d N_{\gamma} }{ \d E }\!
					\int\d M\,
						\Gamma( z )
						\frac{ \d n( M ) }{ \d M }
					\, ,_{_{_{_{_{_{_{_{_{_{_{_{}}}}}}}}}}}}
\end{equation}
where $\Gamma( z )$ is the WIMP annihilation rate around PBHs at redshift $z$ and $\tau_{\Erm}$ is the optical depth back to that redshift. We make the following assumptions.
\begin{itemize}

	\item The spectrum of by-products from WIMP annihilation is obtained using the numerical package in Ref.~\cite{Cirelli:2010xx} \textcolor{blue}{(see also Ref.~\cite{Amoroso:2018qga} for relevant updates)}. Integral~\eqref{eq:flux} only depends on the PBH mass function at $z = 0$, since the $( 1 + z )^{3}$ volume factors cancel out in the computation of Eq.~\eqref{eq:flux}~\cite{Cirelli:2009dv, Cirelli:2010xx}.

	\item The optical depth $\tau_{\Erm}$ in Eq.~\eqref{eq:flux} results from various 
		processes~\cite{Cirelli:2009dv, Slatyer:2009yq}: 
	{\it (i)} photon-matter pair production; 
	{\it (ii)} photon-photon scattering; 
	{\it (iii)} photon-photon pair production.
We adopt the optical depth obtained in Ref.~\cite{Cirelli:2010xx}. We assume that the curvature contribution is zero, in agreement with the prediction from inflation and CMB measurements~\cite{Aghanim:2018eyx, Aghanim:2019ame}, although Refs.~\cite{DiValentino:2019qzk, Handley:2019tkm} argue that the CMB data favours spatially closed models.

	\item We assume a flat FRW metric with the Hubble rate
\begin{equation}
	\qq
	h( z )
		=
					\sqrt{
						\Omega_{\Lambda}
						+
						\Omega_{\mrm}\.( 1 + z )^{3}
						+
						\Omega_{\Rrm}\.( 1 + z )^{4}
					\;}
\end{equation}
in units of $H_{0}$, where we fix the values of the present density parameters to be $\Omega_{\Rrm} = 7 \times 10^{-5}$, $\Omega_{\mrm} = 0.31$ and $\Omega_{\Lambda} = 1 - \Omega_{\mrm} - \Omega_{\Rrm} = 0.69$. The PBH mass function is normalised according to
\begin{equation}
	\label{eq:MF-normalise}
	\int\!\d M\;M\,\frac{ \d n( M,\.z ) }{ \d M }
		 \equiv 
					\rho_{\rm PBH}( z )
					\, ,
\end{equation}
where $\rho_{\rm PBH}( z ) \equiv f_{\rm PBH}\,\rho_{\rm DM}( z )$.
\end{itemize}

Inserting Eq.~\eqref{eq:Gamma} into Eq.~\eqref{eq:flux} and using the normalisation in Eq.~\eqref{eq:MF-normalise}, we can eliminate the mass function in the expression for the flux to obtain
\begin{equation}
	\label{eq:flux-normalised}
	\Phi_{\gamma}
		 = 
					f_{\rm PBH}\,f_{\chi}^{5/3}\.
					\frac{ \Upsilon\.\rho_{\rm DM} }
					{ 2H_{0} }\,
					\tilde{N}_{\gamma}( m_{\chi} )
					\, ,
\end{equation}
where the average number of photons produced is
\begin{equation}
	\label{eq:tildeNgamma}
	\tilde{N}_{\gamma}( m_{\chi} )
		\equiv
					\int_{z_{\star}}^{\infty}\!\d z\;
					\int_{E_{\rm th}}^{m_{\chi}}\!\d E\;
						\frac{ \d N_{\gamma} }{ \d E }
						\frac{ e^{-\tau_{\Erm}(z,\.E)} }
						{ [ h( z ) ]^{1/3} }
					\, .
\end{equation}
The redshift integral is dominated by the range $z \lesssim \Ocal( 100 )$, because of the sharp decline in the optical depth at large redshifts. A numerical fit to the WIMP mass dependence of Eq.~\eqref{eq:tildeNgamma} with the results obtained from the package in Ref.~\cite{Cirelli:2010xx} leads to $\tilde N_{\gamma} \approx 220\.m_{100}^{0.22}$.

Eq.~\eqref{eq:flux-normalised} is valid for all PBH mass distributions, including the monochromatic case~\cite{Boucenna:2017ghj} and the more realistic extended case. For example, one has $\d n / \d M \propto M^{-1/2}$ for PBHs formed from exactly scale-invariant density fluctuations~\cite{Carr:1975qj} or from the collapse of cosmic strings~\cite{Hawking:1987bn} and a lognormal mass function for PBHs formed from a large class of inflationary PBH models~\cite{Green:2016xgy}, such as the axion-curvaton model~\cite{Kawasaki:2012wr}.

Comparing the integrated flux with the Fermi point-source sensitivity $\Phi^{\rm Fermi}_{100\.{\rm MeV}}$ for $f_{\rm PBH} \ll 1$ and $f_{\chi} \approx 1$ yields the limit~\footnote{This constraint could in principle be refined by performing a likelihood analysis accounting for the differential energy spectrum from WIMP annihilation in each energy bin~\cite{Ackermann:2015tah, DiMauro:2015tfa}.}
\begin{equation}
	\label{eq:flux-normalised1}
	f_{\rm PBH}
		\lesssim
					\frac{ 2\.\Phi^{\rm Fermi}_{100\.{\rm MeV}}\,H_{0}}
					{\Upsilon\.
					\rho_{\rm DM}\.
					\tilde{N}_{\gamma}( m_{\chi} ) }
		\approx
					8 \times 10^{-12}\,
					m_{100}^{1.11}
					\, .
\end{equation}
This limit intersects the extragalactic incredulity limit~\eqref{eq:IL} at a mass
\begin{equation}
	\label{eq:MiL}
	M_{\rm IL}^{\rm eg}
		=
					\frac{ 2\.\Phi^{\rm Fermi}_{100\.{\rm MeV}} }
					{ \Upsilon\.H_{0}^{2}\.\tilde N_{\gamma}( m_{\chi} ) }
		\approx
					2.5 \times 10^{10}\,m_{100}^{1.11}\,M_{\odot}
					\, ,
\end{equation}
where the numerical expression accounts for the fit of $\tilde N_{\gamma}( m_{\chi} )$ and we set $\sv = \sv_{\Frm}$. This corresponds to an upper limit on the mass of a SLAB in our Universe.

\subsection{Flux from nearest individual source}

The gamma-ray background flux produced from dark-matter annihilation around an individual PBH is~\cite{Bringmann:2011ut}
\begin{equation}
	\label{eq:flux0}
	\frac{ \d\Phi_{\gamma} }{ \d E\.\d\Omega }
		 = 
					\frac{ \Gamma }{ 8\pi\.d_{\Lrm}^{2} }\.
					\frac{ \d N_{\gamma} }{ \d E }
					\, ,
\end{equation}
where $\Gamma$ is the DM decay rate around the BH and $d_{\Lrm}$ is the distance of the PBH, which is necessarily extragalactic in the SLAB case. The BH-halo system can be detected within a distance
\begin{equation}
	\label{eq:effectived}
	d_{\Lrm}
		 = 
					\sqrt{\,
						\frac{ \Gamma\,N_{\gamma}( m_{\chi} ) }
						{ 2\,\Phi^{\rm Fermi}_{100\.{\rm MeV}} }
					\,}
					\, ,
\end{equation}
where the average number of photons resulting from the annihilation processes is
\begin{equation}
	\label{eq:Ngamma}
	N_{\gamma}( m_{\chi} )
		 = 
					\int_{E_{\rm th}}^{m_{\chi}}\!\d E\;
					\frac{ \d N_{\gamma} }{ \d E }
					\, .
\end{equation}
We fit the numerical solution of Eq.~\eqref{eq:Ngamma} with the package in Ref.~\cite{Cirelli:2010xx} to obtain 
\begin{equation}
	N_{\gamma}
		\approx
					2.0\.( m_{\chi} / {\rm GeV} )^{0.32}
					\, .
\end{equation}
The ratio $\tilde{N}_{\gamma}( m_{\chi} ) / N_{\gamma}( m_{\chi} )$ can be estimated analytically by neglecting the $E$-dependence of the opacity and assuming $h( z ) \approx \Omega_{\mrm}^{1/2}( 1 + z )^{3/2}$. Then Eq.~\eqref{eq:tildeNgamma} implies $\tilde{N}_{\gamma} \approx 50\.N_{\gamma}$.

\begin{figure}[tb]
	\vs{4mm}
	\includegraphics[width = 0.99\linewidth]{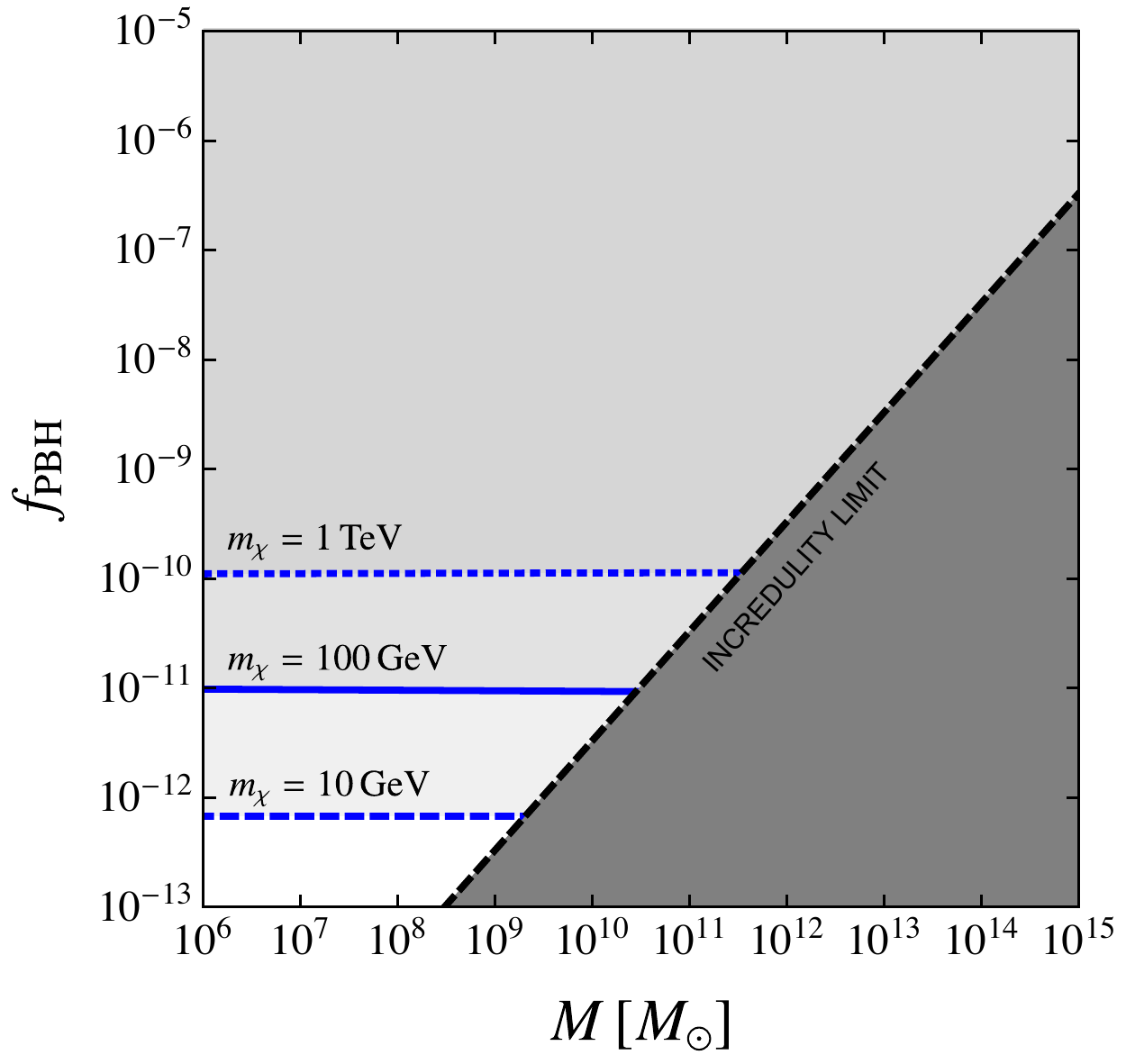} 
	\caption{Constraints on $f_{\rm PBH}$ as a function of PBH mass.
 			Results are shown for 
			$m_{\chi} = 10{\rm\,GeV} / c^{2}$ (dashed line), 
			$m_{\chi} = 100{\rm\,GeV} / c^{2}$ (solid line) and 
			$m_{\chi} = 1{\rm\,TeV} / c^{2}$ (dotted line), 
			setting $\sv = 3 \times 10^{-26}\,$cm$^{3}$/s.
			Also shown is the incredulity limit (black dashed line)}
	\label{fig:fPBH}
\end{figure}

For a given value of $M$, we can compare Eq.~\eqref{eq:effectived} with the expected distance to the nearest BH, $d \approx ( M / \rho_{\rm BH} )^{1/3}$. If BHs are primordial, $\rho_{\rm BH} = f_{\rm PBH}\.\rho_{\rm DM}$, so this leads to the constraint 
\begin{equation}
	\label{eq:limit}
	f_{\rm PBH}
		\lesssim
					\!\left(
						\frac{ 2\.\Phi^{\rm Fermi}_{100\.{\rm MeV}} }						
						{\Upsilon\.N_{\gamma}( m_{\chi} ) }
					\right)^{\!\!3/2}\!
					\frac{ M^{-1/2} }{ \rho_{\rm DM} }
					\, .
\end{equation}
The BH is necessarily extragalactic in the SLAB case and this analysis holds providing it is at a redshift $z \ll 1$. The bound~\eqref{eq:limit} is more stringent than the background bound~\eqref{eq:flux-normalised1} only if $M$ exceeds
\begin{equation}
	\label{eq:barM}
	\bar M
		\equiv
					\frac{ 2\.\Phi^{\rm Fermi}_{100\.{\rm MeV}} }
					{ \Upsilon\.H_{0}^{2} }\,
					\frac{ \tilde N_{\gamma}^{2}( m_{\chi} ) }
					{ N_{\gamma}^{3}( m_{\chi} ) }
		=
					M_{\rm IL}^{\rm eg}\mspace{-2mu}
					\[
						\frac{\tilde N_{\gamma}( m_{\chi} ) }
						{ N_{\gamma}( m_{\chi} ) }
					\]^{3}
					,
\end{equation}
where $M_{\rm IL}^{\rm eg}$ is given in Eq.~\eqref{eq:MiL}. However, the quantity in square brackets is $\Ocal( 10^{5} )$, so $\bar M \gg M_{\rm IL}^{\rm eg}$ and the nearest-source bound lies well outside the cosmological incredulity limit. Therefore the individual bound is never applicable for the range of WIMP masses considered. Figure~\ref{fig:fPBH} shows the constraints on $\fPBH$ for different WIMP masses: $m_{\chi} = 10\,$GeV (dashed lines), $m_{\chi} = 100\,$GeV (solid lines), $m_{\chi} = 1\,$TeV (dotted lines). We extend the computation to a wider range of WIMP and BH masses in a follow-up paper~\cite{Visinelli:2020B}.

\section{SLABs and Light Bosons}
\label{sec:SLABs-and-Light-Bosons}

Is it possible for spinning BHs to lose a portion of their rotational energy via the interaction with an interfering boson wave of frequency $\omega < \mu\,\Omega_{\rm BH}$, where $\Omega_{\rm BH}$ is the BH horizon frequency and $\mu$ is the azimuthal number of the wave. When this criterion is satisfied, the outgoing wave extracts energy and angular momentum from the BH through the phenomenon of superradiance. If light bosonic fields exist in nature, they could accumulate around rotating SMBHs and form a condensate, leading to such superradiant instabilities~\cite{Press:1972zz}. A portion of the rotational energy of the SLAB might be dissipated by the boson cloud via superradiance if the Compton wavelength of the boson $\lambda_{\Crm} = h / ( m_{\phi} c )$, where $h$ is the Planck constant and $m_{\phi}$ is its mass, is comparable to the Schwarzschild radius of the SLAB. Interestingly, for $M \gtrsim 10^{10}\,M_{\odot}$, this condition is realised for an ultra-light boson of mass $m_{\phi} \lesssim 10^{-22}\,$eV, which has important astrophysical consequences~\cite{Hu:2000ke, Visinelli:2018utg}. For example, this mechanism has been applied jointly with the observations of the mass and spin of the SMBH M87$^{*}$ to place bounds on the mass of hypothetical light bosons~\cite{Davoudiasl:2019nlo}. Although PBHs are generally formed with a negligible spin, we expect SLABs to acquire a large momentum due to the accretion mechanisms described in Sec.~\ref{sec:Accretion-Constraints-on-SLABs}. We discuss the phenomenon for light spin-zero fields, although important consequences are also obtained for spin-one fields~\cite{Pani:2012vp, East:2017ovw} and tensor fields~\cite{Brito:2015oca}.

The dimensionless spin parameter is $a_{*} = J c / ( G M^{2} )$, where $J$ is the angular momentum of the SLAB. When the angular velocity of the BH horizon is larger than the angular phase velocity $\omega$ of the wave, a population of spin-zero bosons grows around a spinning BH~\cite{Arvanitaki:2009fg, Arvanitaki:2010sy},
\begin{equation}
	\label{eq:BHsuperradiance}
	\omega
		<
					\frac{ \mu }{ r_{\Srm} }\,
					\frac{ a_{*} }{ 1 + \sqrt{1 - a_{*}^{2}\,} }
					\, .
\end{equation}
The square root term ensures that the spin parameter of a Kerr BH cannot exceed unity.

The leading mode of the superradiant bound state of scalar bosons grows exponentially, $N_{\mu} \propto \exp( \Gamma_{\phi} t )$, at a rate~\cite{Baryakhtar:2017ngi} 
\begin{equation}
	\Gamma_{\phi}
		=
					a_{*}\.r_{\grm}^{8}\.
					m_{\phi}^{9} / 24
					\, .
\end{equation}
For example, for an ultra-light axion of mass $m_{\phi} = 10^{-22}\,$eV and with a Compton wavelength comparable with $r_{\grm}$, the rate is $\Gamma_{\phi} \approx 10^{-8}\,\srm^{-1}$.

Superradiance is disrupted over a characteristic BH timescale $\tau_{\rm BH}$, related to the accretion timescale by
\begin{equation}
	\label{eq:BHdepletion}
	\Gamma^{}_{\phi}\.\tau_{\rm BH}
		\gtrsim
					\ln N_{\mu}
					\, .
\end{equation}
We take the characteristic BH timescale to be $\tau_{\rm BH} \sim t_{\Srm}$~\cite{Baryakhtar:2017ngi}, where $t_{\Srm}$ is the Salpeter timescale introduced in Sec.~\ref{sec:Formation-of-SLABs} and we use an efficiency parameter $\epsilon \sim 0.1$~\cite{Shankar:2007zg}. The occupation number of the boson cloud for the azimuthal number $\mu$ after the SLAB has spun down by a value $\Delta a_{*}$ is~\cite{East:2017ovw}
\begin{equation}
	N_{\mu}
		=
					\frac{ G M^{2} \Delta a_{*} }{ \mu }
					\, .
\end{equation}
 
\begin{figure}[tb]
	\vs{4mm}
	\includegraphics[width = 0.99\linewidth]{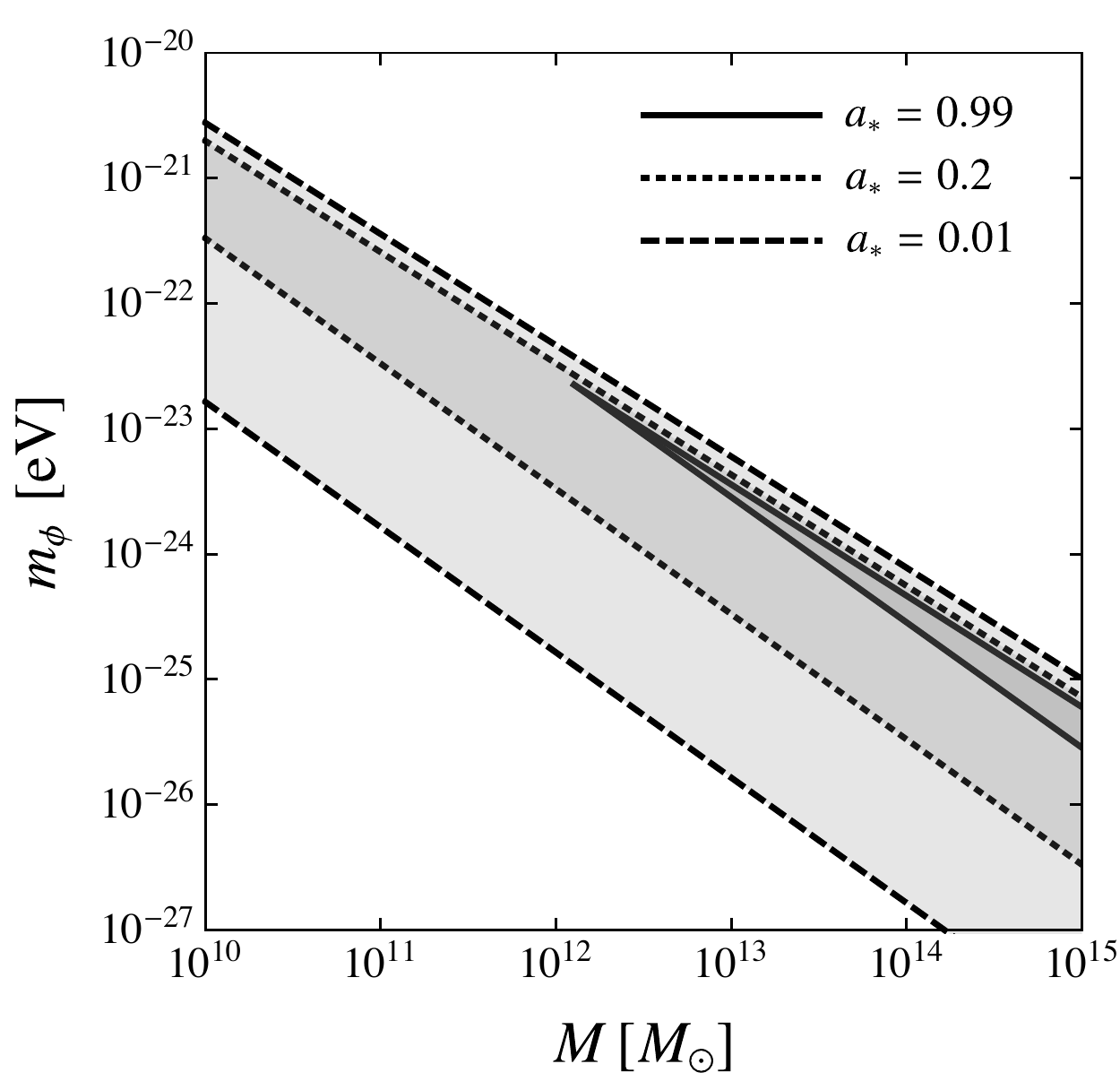} 
	\caption{Superradiance constraints on the mass $m_{\phi}$ of a hypothetical boson 
			as a function of BH mass $M$.
			Results are shown for the observed black-hole spin 
				$a_{*} = 0.99$ (solid line), 
				$a_{*} = 0.2$ (dashed line) and 
				$a_{*} = 0.01$ (dotted line).}
	\label{fig:BHboson}
\end{figure}

If a BH with spin $a_{*}$ is observed, the condition in Eq.~\eqref{eq:BHsuperradiance} yields a {\it lower} bound on the mass of the light boson, $m_{\phi} \approx \omega$, while the requirement of Eq.~\eqref{eq:BHdepletion} that superradiance has not depleted the spin of the BH by the amount $\Delta a_{*}$ leads to an {\it upper} bound on $m_{\phi}$. Figure~\ref{fig:BHboson} shows the region excluded by Eqs.~\eqref{eq:BHsuperradiance} and~\eqref{eq:BHdepletion}. We assume that the SLAB had an initial spin $a_{*\.i} \approx 1$ and evolved so that its spin today is $0.99$ (solid line), $0.2$ (dotted line) or $0.01$ (dashed line). Since the Schwarzschild radius of a SLAB is considerably larger than that of M87$^{*}$, observing these objects would lead to a constraint for extremely light bosons with mass $m_{\phi} \ll 10^{-20}\,$eV.

\section{Results and Discussion}
\label{sec:Results-and-Discussion}

In this work, we have examined the bounds on stupendously large BHs with $M \gtrsim 10^{11}\,M_{\odot}$, here referred to as SLABs. We have considered their possible formation mechanisms and assessed the limits coming from dynamical, lensing and accretion effects and from gamma-ray annihilation of WIMP dark-matter around PBHs.

We have assumed that the WIMP cross-section does not change during the evolution of the Universe. This is not true if there is a light mediator that leads to a Sommerfeld enhancement of the WIMP annihilation~\cite{ArkaniHamed:2008qn}. We have also assumed that the cross-section is fixed to the value obtained at freeze-out $\sv_{\Frm}$ in the standard cosmological model. However, its value might deviate considerably from this if there were an early period in which the cosmological density was dominated by matter or some other exotic form of energy~\cite{Gelmini:2008sh}. The expected signal from WIMPs annihilating around a SLAB also needs to be reconsidered if the WIMP velocity distribution plays a r{\^o}le in the computation of $\sv$, for example when corrections of order $( v / c )^{2}$ are to be taken into account or when the annihilation does not proceed through an $s$-channel.

The expected gamma-ray flux from WIMP annihilation depends on the combination $\Phi_{\gamma} \propto f_{\rm PBH}\,f_{\chi}^{5/3}$, as shown by Eq.~\eqref{eq:flux-normalised}. In this work, we have assumed that WIMPs make up most of the DM, with PBHs contributing a negligible fraction. However, this reasoning can be inverted to constrain the WIMP fraction $f_{\chi}$ when PBHs form most the DM. We explore the consequences of this in an accompanying paper~\cite{Visinelli:2020B}.

In Sec.~\ref{sec:SLABs-and-Light-Bosons} we have discussed the possible constraints on the mass of ultra-light bosons for a given SLAB spin due to superradiance effects. Although SMBHs nearly as large as SLABs are known to exist, they are considerably further away than M87$^{*}$ or Sagittarius A$^{*}$, making the determination of their spin and their imaging more challenging~\cite{Shemmer:2004ph}. Furthermore, their accretion effects would modify the size of the black-hole shadow with time. We leave these consideration for future work.

Our discussion has not covered other possible particle DM candidates, like the sterile neutrino~\cite{Boyarsky:2018tvu} or the QCD axion~\cite{DiLuzio:2020wdo}. If SLABs are present in the Universe, they would provide a powerful tool for cosmological tests due to their unique imprints. In fact, the stringent bound on the fraction of PBHs given by Eq.~\eqref{eq:limit} is based on this assumption. However, our constraints cannot be applied to models in which PBHs provide nearly all the DM.

\acknowledgments
We thank Niayesh Afshordi, Yacine Ali-Ha{\"i}moud, Ben Horowitz, Priya Natarajan, Rafaek Nunes, Alex Vilenkin and Kumar Shwetketu Virbhadra and for helpful comments. F.K.~Acknowledges hospitality and support from the Delta Institute for Theoretical Physics. L.V.~acknowledges support from the NWO Physics Vrij Programme ``The Hidden Universe of Weakly Interacting Particles'' with project number 680.92.18.03 (NWO Vrije Programma), which is (partly) financed by the Dutch Research Council (NWO).

\setlength{\bibsep}{5pt}
\setstretch{1}
\bibliography{PBHbib}

\end{document}